\documentclass[useAMS,usenatbib,usegraphicx]{mn2e}

\usepackage{aas_macros}
\usepackage{amssymb}
\usepackage{hyperref}

\newcommand{\Cs}{{\bf C}}
\newcommand{\Os}{{\bf O}}
\newcommand{\SB}{{\rm SB}}
\newcommand{\obs}{{\rm obs}}

\begin{document}

\author[Hans F. Stabenau, Andrew Connolly, Bhuvnesh Jain]{
  Hans F. Stabenau$^1$\thanks{E-mail: hstabena@astro.upenn.edu},
  Andrew Connolly$^2$\thanks{E-mail: ajc@astro.washington.edu},
  Bhuvnesh Jain$^1$\thanks{E-mail: bjain@astro.upenn.edu} \\
$^1$University of Pennsylvania, Department of Physics and Astronomy, Philadelphia, PA 19104, USA\\
$^2$University of Washington, Astronomy Department, Seattle, WA 98195, USA}

\title[Photo-z's with surface brightness priors]{Photometric redshifts with surface brightness priors}

\date{Submitted 2007 December 10}

\maketitle

\begin{abstract}
  We use galaxy surface brightness as prior information to improve
  photometric redshift estimation. We apply our template-based photo-z
  method to imaging data from the ground-based VVDS survey and
  the space-based GOODS field from HST, and use spectroscopic redshifts to test
  our photo-z's for different galaxy types and redshifts. We find that
  the surface brightness prior eliminates a large fraction of outliers
  by lifting the degeneracy between the Lyman and 4000~\AA ngstrom
  breaks.  
  Bias and scatter are improved by about a factor of two with the
  prior
  in each redshift bin in the range $0.4<z<1.3$, for both the ground
  and space data.  Ongoing and planned surveys from the ground and
  space will benefit provided that care is taken in measurements of
  galaxy sizes and in the application of the prior. We discuss the
  image quality and signal-to-noise requirements that enable the
  surface brightness prior to be successfully applied.
\end{abstract}


\section{Introduction}
\label{sec:intro}

Photometric redshifts (``photo-z's'') are a tool for obtaining
redshift and type information for galaxies for which broadband
photometric colors are available rather than spectroscopic data.
Generally, this technique is used to study populations of galaxies
when the observational cost of obtaining spectra for these galaxies
would be prohibitive. With the advent of wide field imaging surveys,
photo-z estimation has become an indispensable part of cosmological
surveys. With imaging in three to five optical bands, ongoing and
planned surveys aim to get photo-z's over hundreds or thousands of
square degrees of sky.  With the redshift and sky positions, 
  surveys such as
  DES\footnote{\url{http://www.darkenergysurvey.org/}},
  KiDS\footnote{\url{http://www.strw.leidenuniv.nl/~kuijken/KIDS/}},
  LSST\footnote{\url{http://www.lsst.org/}}, and
  Pan-STARRS\footnote{\url{http://pan-starrs.ifa.hawaii.edu/}}
  plan to use galaxy
  clusters, galaxy clustering and gravitational lensing as probes of
  dark energy and other cosmological issues. The techniques and
accuracy of photo-z's have become an area of active study since they
play a critical role in deriving cosmological information from imaging
surveys.


Recently, two main approaches have been used to obtain photo-z's:
empirical fitting and template fitting.  In the former, a neural
network, polynomial function, or other empirical relation is trained
using a subsample of galaxies with known (spectroscopic) redshifts,
and then applied to the larger sample \citep{Connolly:1995yq,
  Vanzella:2003ca, Firth:2002yz, Brodwin:2006hh, Collister:2006qg,
  Oyaizu:2007kd, Raffaele:2007te, Li:2006vd, 
  Abdalla:2007uc,Banerji:2007nh}.  In the template-fitting method, first a
library of theoretical or empirical spectral energy distributions
(``SEDs'') are generated, and then they are fit to the observed colors
of galaxies, where the redshift is a parameter that is fit
\citep{Arnouts:1999bb, Benitez:1998br, 2000A&A...363..476B,
  Babbedge:2004iz, Ilbert:2006dp, Brodwin:2006hh, Feldmann:2006wg,
  Mobasher:2006fj, 2006ApJS..162...20B, Margoniner:2007qg, Wittman:2007hy}. 
As in the training set
approach, a subsample of galaxies with spectroscopic redshifts can be
used to help calibrate this procedure.  In this study we use a
template-fitting method, which makes use of a library of galaxy
templates \citep{1993ApJ...405..538B, Bruzual:2003tq}, to predict the
colors of galaxies through a series of optical and infrared passbands.
In principle the idea is simple: given an observed galaxy, compute the
redshift $z$ that makes each template $T$ match the observed colors
most closely, and then choose the best fit $(T,z)$ pair.

The reality is more difficult.  With only a limited number of colors,
it's often impossible to tell whether a set of observed colors better
matches one template which is at high $z$, or another at low $z$.
This is known as the ``color-redshift degeneracy,'' and occurs because
a red galaxy template at low redshift and a blue galaxy template at
high redshift can look the same in a given set of filters.  For the
same reason, accurately determining the rest-frame color of a galaxy
(the ``$k$-correction'') is difficult: intrinsic astrophysical
variations and evolution broaden scatter in the color-$z$ relation,
and the effect increases at higher redshift. With this paper we
address the question of how to break this degeneracy. The most widely
used method uses empirically measured apparent magnitude--redshift
distributions \citep[see e.g.][]{Benitez:1998br}; in this paper, we
introduce a new and complementary approach to breaking the
color-redshift degeneracy using the surface brightness (SB), which is
the luminosity of an object per unit surface area.  We will show that
the SB is able to provide a strong constraint on galaxy redshifts,
which should remain intact even for faint samples, and that this will
not be the case for the magnitude prior.  In order to incorporate
prior statistical information about the galaxies under study into the
redshift estimate in a consistent way, we use the Bayesian approach
first introduced by \citet{Benitez:1998br}.

Throughout this paper we use the AB magnitude system.

\begin{figure*}
  \centering
  \includegraphics[width=0.45\linewidth]{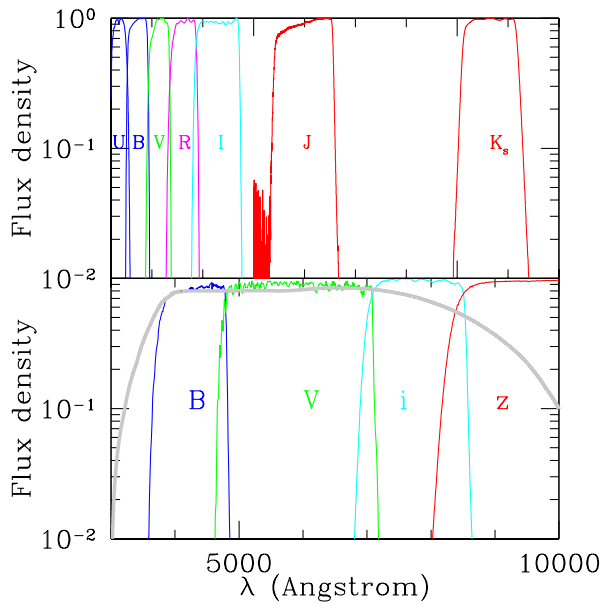}
  \caption{
    Flux density vs wavelength for the
    filters used in the GOODS and VVDS surveys.  The ACS filters used
    are F435W (Johnson $B$), F606W (Broad $V$), F775W (SDSS $i$), and F850LP
    (SDSS $z$).  The VVDS filters are Mould $U$, $B$, $V$, $R$, and $I$
    filters, as well as ESO/WFI $J$ and $K_s$ filters.  The detector
    response function for the ACS is indicated on the lower plot.}
  \label{fig:filters}
\end{figure*}

\section{Bayesian Photometric Redshifts}
\label{sec:bayes}

Our goal is to find, for each galaxy, the 2-D posterior distribution
$P(z,T|\Cs,\Os)$, where $z$ is the redshift of the galaxy, $T$ is the
``template parameter,'' which is a discrete variable corresponding to
the galaxy type in our template library, $\Cs$ is the vector of fluxes
from the data, and $\Os$ is a vector of observables independent of the
fluxes $\Cs$ for each galaxy.  $\Os$ could be any set of observables,
e.g.\ size, brightness, morphology, environment. In the current paper
we consider $I$-band apparent magnitude and apparent surface
brightness.  Using Bayes' theorem and the definition of conditional
probability, we can write
\begin{equation}
  \label{eq:pz1}
  P(z,T|\Cs,\Os) = \frac{P(\Cs,\Os,z,T)}{P(\Cs,\Os)} 
  = \frac{P(\Cs|\Os,z,T)P(\Os,z,T)}{P(\Os)P(\Cs)},
\end{equation}
because we make the approximation that $\Os$ and $\Cs$ are
independent.  By the same token, we can write $P(\Cs|\Os,z,T) =
P(\Cs|z,T)$, and then since $P(z,T,\Os)/P(\Os) = P(z,T|\Os),$
\begin{equation}
  \label{eq:pz2}
  P(z,T|\Cs,\Os) = \frac{P(\Cs|z,T)P(z,T|\Os)}{P(\Cs)}.
\end{equation}
This is the equation on which our work is based.  The posterior
distribution $P(z,T|\Cs,\Os)$ is given in terms of the likelihood
function $P(\Cs|z,T)$ and the prior distribution $P(z,T|\Os)$; the
prior encompasses all the prior knowledge we wish to use about galaxy
morphology, evolution, environment, brightness, etc.  Once the
posterior distribution is determined, all the quantities of interest
can be calculated; ideally, any study that wishes to make use of
photo-z information would use the full $P(z,T|\Cs,\Os)$ directly.  For
the sake of simplicity, we evaluate the performance of our estimator
using the mode of $P(z,T|\Cs,\Os)$, which is the ``best'' redshift
\begin{equation}
  \label{eq:zp}
  z_p \equiv z(P=P_{\rm max}).
\end{equation}

We can also compute the marginalized redshift estimate given by
\begin{equation}
  \label{eq:marg}
  \langle z \rangle = \frac{\int dT dz \, z P(z,T|\Cs,\Os)}
  {\int dT dz \, P(z,T|\Cs,\Os)},
\end{equation}
and the variance
\begin{equation}
  \label{eq:variance}
  \langle z^2 \rangle - \langle z \rangle^2
  = \frac{\int dT dz \, z^2 P(z,T|\Cs,\Os)}
  {\int dT dz \, P(z,T|\Cs,\Os)} - \langle z \rangle^2.
\end{equation}
As we will show, we find that $P(z,T|\Cs)$ is typically multimodal
with one minimum in $\chi^2$ close to the spectroscopic redshift and
one or more ``false'' minima at other redshifts.  Using $z_p$ from
Eq.~(\ref{eq:zp}) as the redshift estimator means that, without a
prior, the algorithm can often generate a large error by choosing the
estimate from one of the false minima.  Including a prior deweights
those minima and reduces the impact of the outliers (often referred to
as ``catastrophic'' outliers), greatly improving the accuracy of the $z_p$
estimate.  Ideally, however, any study using photo-z information
should use the full posterior distribution $P(z,T|\Cs,\Os)$ in its
analysis.

\section{Template Fitting}
\label{sec:templates}


\begin{figure*}
  \centering
  \includegraphics[width=0.75\linewidth]{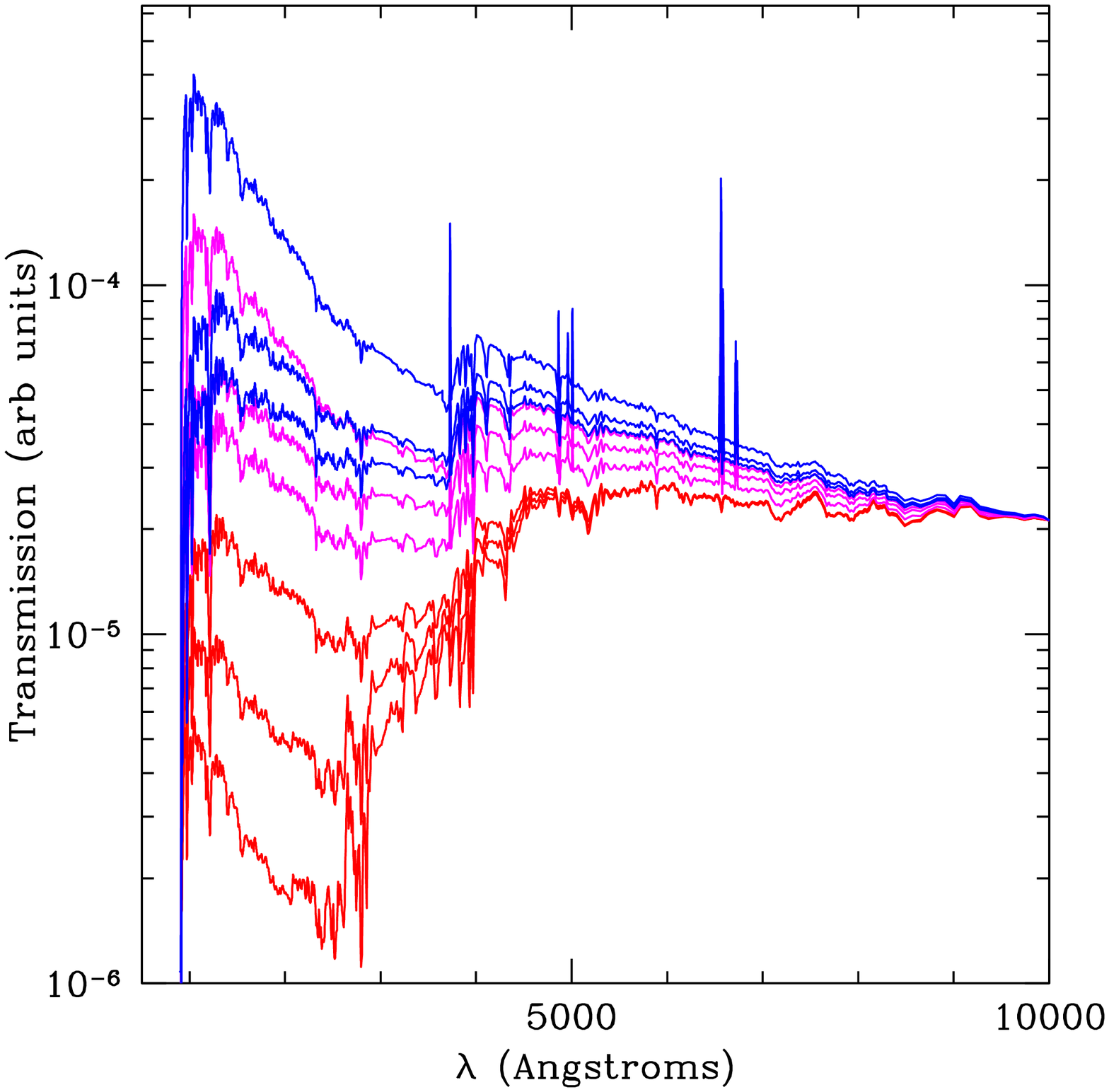}
  \caption{
    Galaxy templates \citep[GISSEL][]{1993ApJ...405..538B,
      Bruzual:2003tq} transmission vs $\lambda$, overplotted and
    normalized to the same value at 10000~\AA.  The templates on the
    bottom, shown in red, are the reddest and oldest galaxy
    populations, while those toward the top are bluer and younger; the
    ages range from 14~Gyr ($T=0$) to 50~Myr ($T=9$).  Convolution
    with the survey filters flattens out the emission lines, while the
    4000~\AA\ and Lyman breaks are still pronounced, which are the
    primary features of the galaxy spectra that constrain photo-z's.}
  \label{fig:templates}
\end{figure*}

In order to calculate the likelihood function $P(\Cs|z,T)$, we
initially model the colors of galaxies as a function of redshift. We
use the simple stellar population models from the GISSEL (Galaxy
Isochrone Synthesis Spectral Evolution Library) spectral synthesis
package \citep{1993ApJ...405..538B, Bruzual:2003tq} to derive a set of
spectral energy distributions with ages from 50 Myr to 15 Gyrs.  We
parameterize the galaxy type as a one dimensional sequence, $T$, that
expresses the age of the spectrum with $T=9$ for galaxies with ages of
50 Myr through to $T=0$ for galaxies of age 15 Gyr.
All SEDs are convolved with a set of filter response functions taken
from the GOODS \citep{Giavalisco:2003ig} and VVDS
\citep{LeFevre:2003mf} surveys to model the color as a function of
redshift over the redshift interval $0<z<4$ (see
Figs.~\ref{fig:filters} and \ref{fig:templates}). All input spectra
are corrected for attenuation due to the intergalactic medium
\citep{1995ApJ...441...18M}. We make no account for evolution in the
SEDs as a function of redshift nor do we train the spectral models to
correct for uncertainties within the photometric zero-points of the
observed data; 
instead we choose a largely equivalent, and simpler, procedure of fitting a
correction to the bias on the calibration sample for use on the
(separate) test sample.


Given a model of galaxy colors as a function of redshift and spectral
type, $G_i(z,T)$, whose values are the flux in the $i^{\rm th}$ band
for a given template $T$ and redshift $z$ we then calculate, for each
galaxy,
\begin{equation}
  \label{eq:gal_chisq}
  \chi^2(z,T) = \sum_i (C_i - \alpha G_i(z,T))^2/\sigma_i^2,
\end{equation}
where $P(\Cs|z,T) \equiv \exp(-\chi^2(z,T)/2)$, $C_i$ are the observed
fluxes, $\sigma_i$ the photometric uncertainties and $\alpha$ is a
scale factor that is a free parameter for each galaxy. The scale
factor, $\alpha$, means that the template fitting procedure by itself
does not take into account the overall brightness of a galaxy, but
only its colors.  Since we do not have an accurate way to estimate the
true error in our template fitting procedure, once we have calculated
the full 2-D likelihood function $P(\Cs|z,T)$, we scale the errors on
the measured $\Cs$ so that $\chi_{\rm min}(z,T) = 1$. Once we have
determined the prior distribution $P(z,T|\Os)$, we merely have to
multiply the two distributions to get the desired posterior
distribution.  We address how the priors are generated in detail in
the next section.  

For the current work we utilize the galaxy catalogs from the GOODS and
VVDS redshift surveys. For the GOODS data we utilize the $B,V,i,z$
photometric passbands and for the VVDS the $U,B,V,R,I,J,K_s$ passbands
(see Fig.~\ref{fig:filters}).

\section{Priors}
\label{sec:priors}

\begin{figure*}
  \centering
  \includegraphics[width=0.45\linewidth]{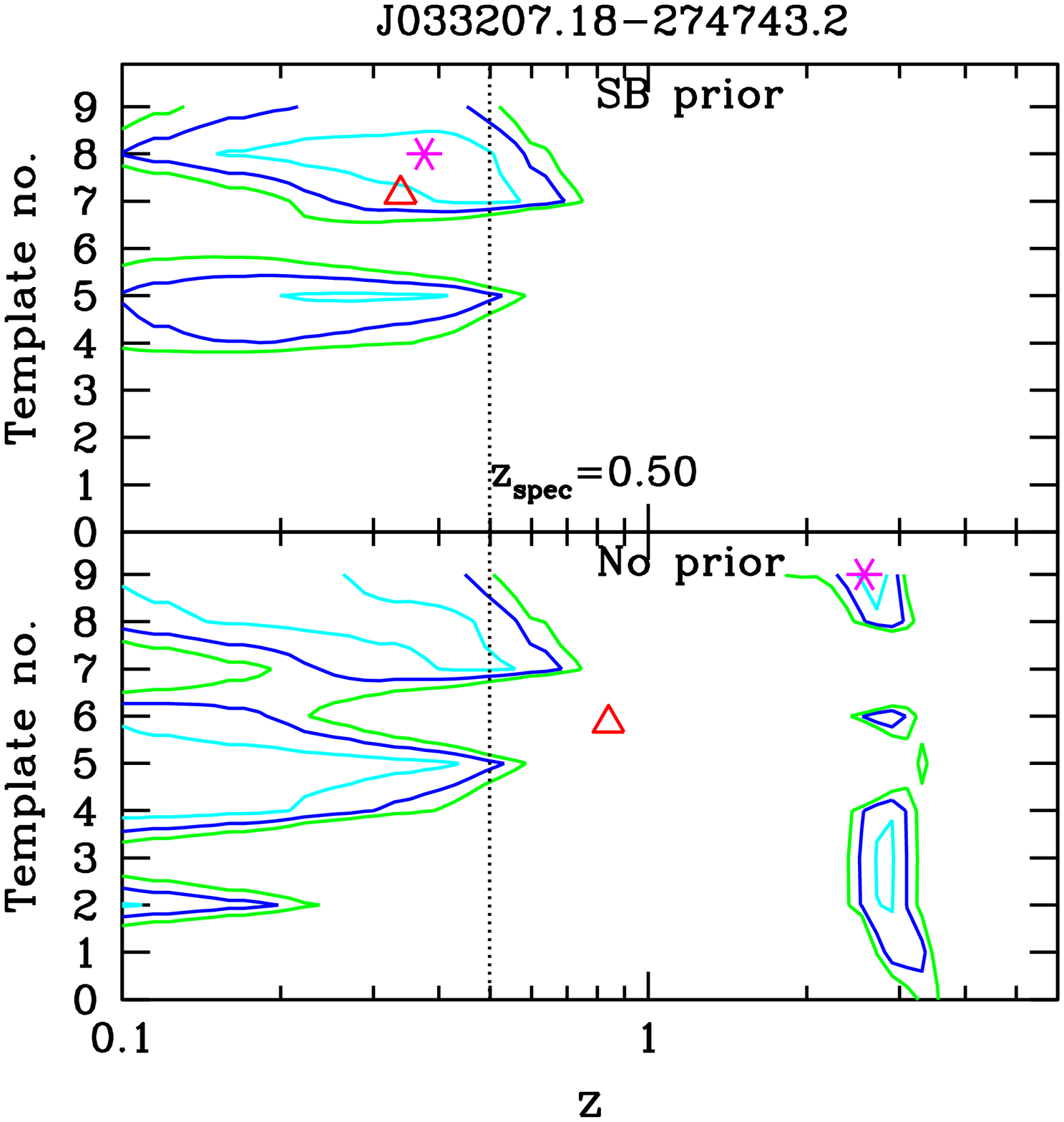}
  \includegraphics[width=0.45\linewidth]{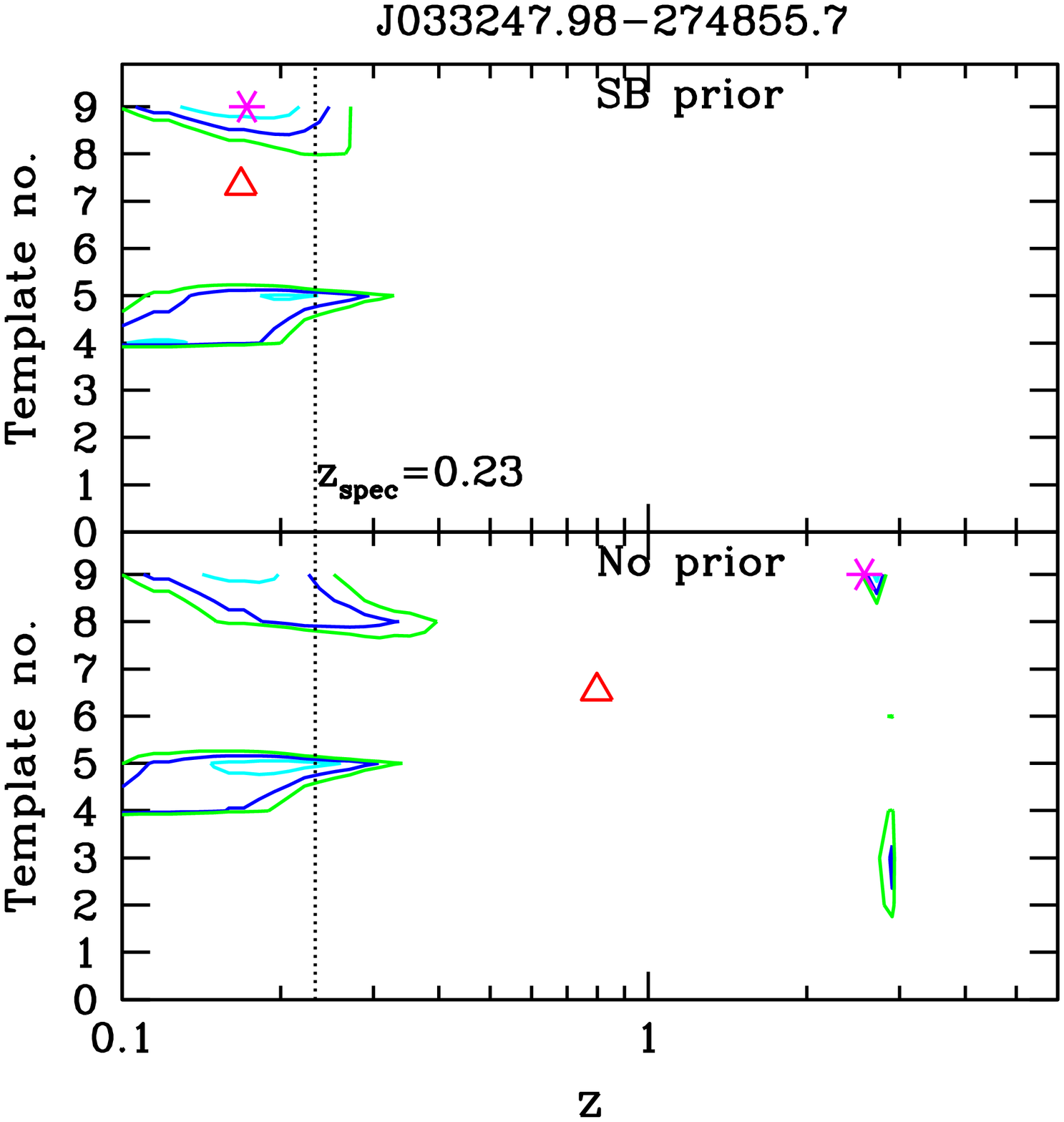}
  \caption{
    The two panels contain $1-$ (cyan), $2-$ (blue), and
    $3-\sigma$ (green) probability contours for the two galaxies in
    Fig.~\ref{fig:galspectra}.  Each galaxy has a multiply-peaked
    $P(z,T)$ distribution where the template fitting approach fails to
    select the correct peak without a prior; the red triangle and
    magenta asterisk marks denote the values of $\langle z \rangle$
    and $z(\chi=\chi_{\rm min})$, respectively.  For each galaxy, the
    upper panel shows how the SB prior breaks the degeneracy by
    eliminating the inaccurate high-redshift maximum, moving both
    $\langle z \rangle$ and $z_{\rm min}$ closer to the measured
    spectroscopic redshift, denoted by the dashed line.  }
  \label{fig:degengals}
\end{figure*}

\begin{figure*}
  \centering
  \includegraphics[width=0.45\linewidth]{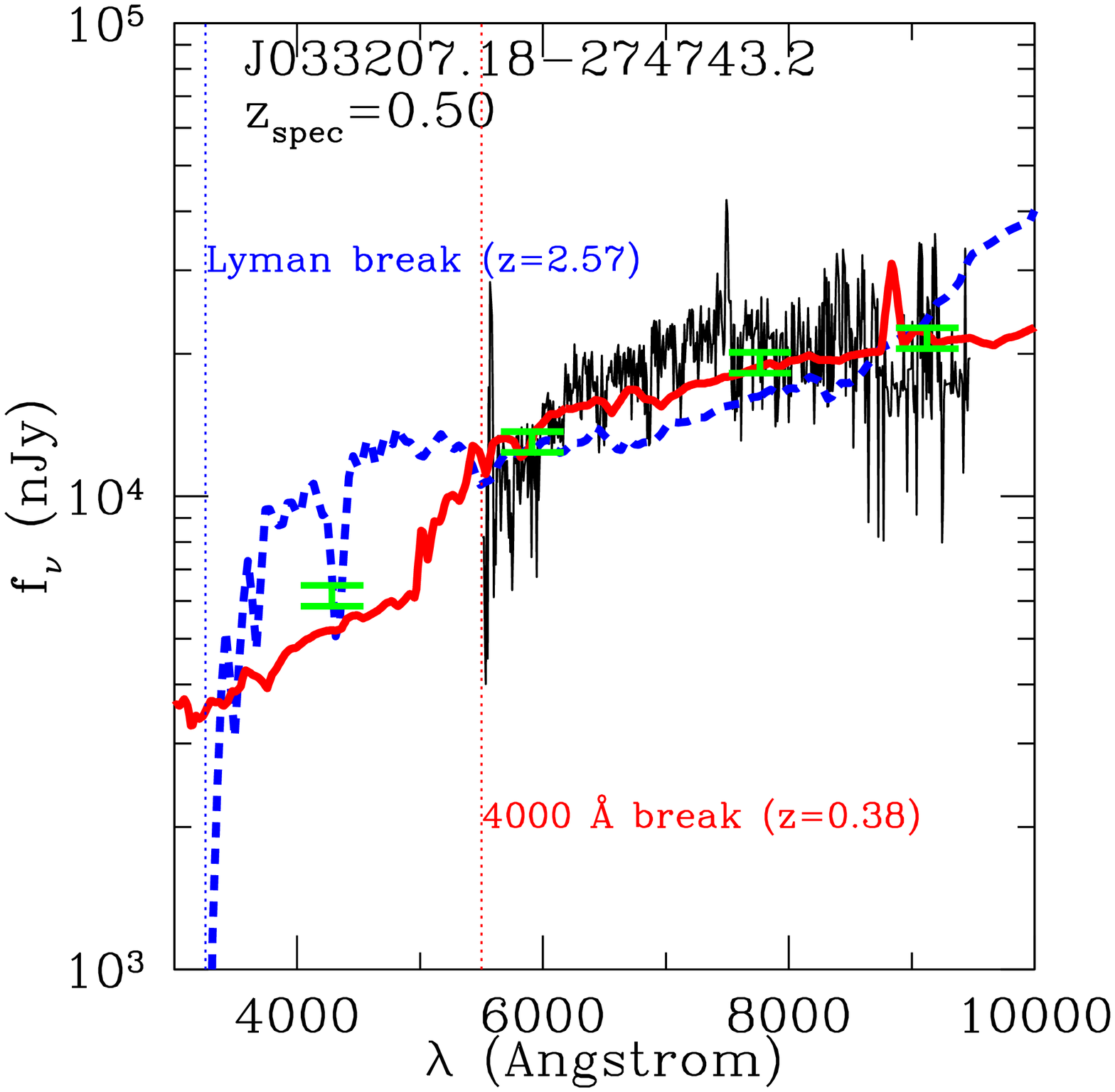}
  \includegraphics[width=0.45\linewidth]{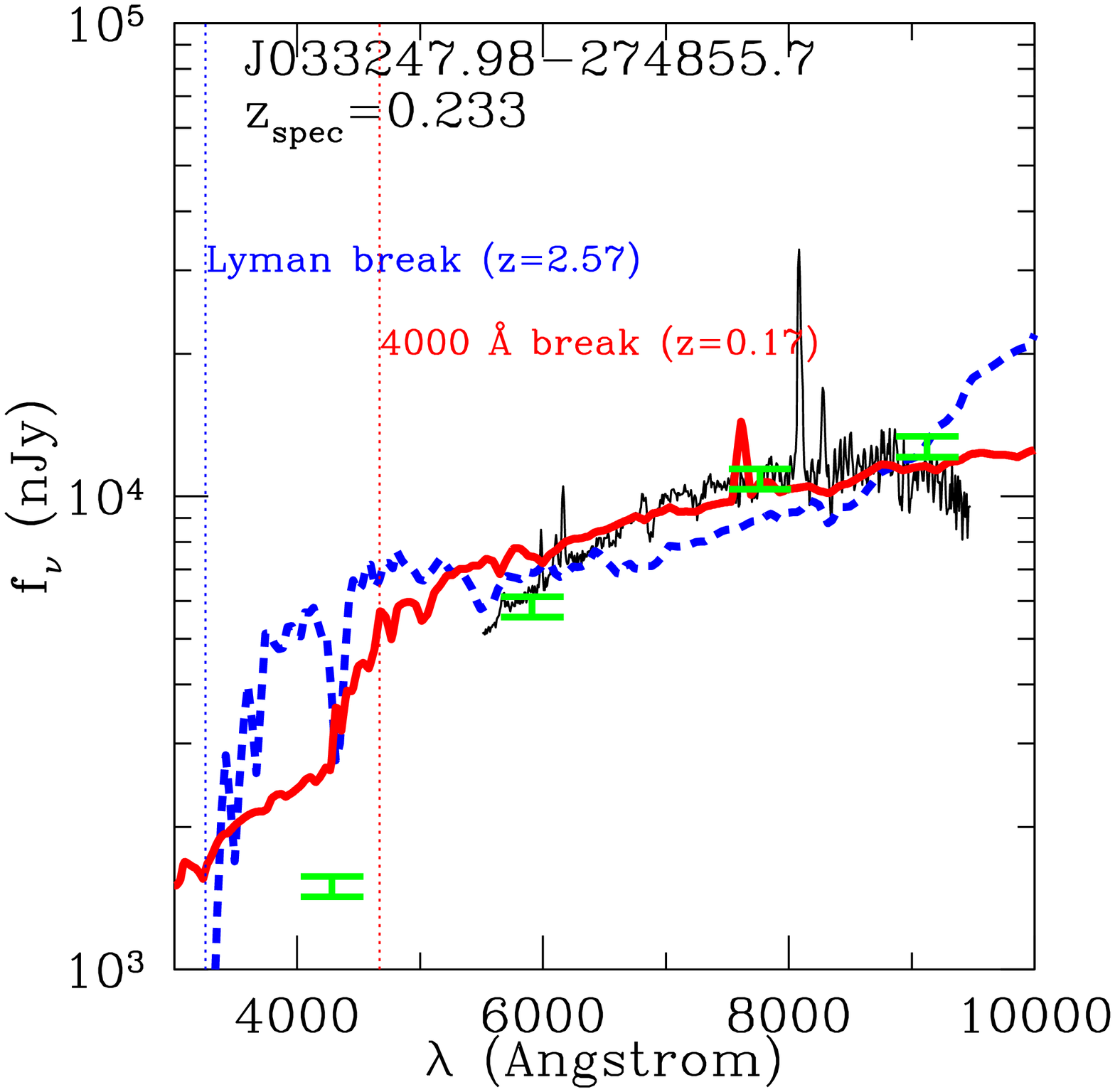}
  \caption{
    Flux vs wavelength for the two degenerate galaxies shown in
    Fig.~\ref{fig:degengals}.  The black curve is the high-resolution
    spectrum of the galaxy, the dashed blue and solid red curves are
    the template fits at high and low redshift, respectively, and the
    green bars are the observations in the broad band filters.  The
    Lyman and 4000~\AA\ breaks are shown by dashed lines.  For each
    galaxy, a $P(z,T)$ distribution is computed by fitting the
    templates to the observed galaxy colors.  However, without prior
    information, it is nearly impossible to tell which break is
    present in the spectrum, and therefore, whether the galaxy is at
    high or low redshift.}
  \label{fig:galspectra}
\end{figure*}

The choice of prior $P(z,T|\Os)$ is key to a successful template-based
photo-z study.  The typical failure mode for a template-based photo-z
code comes in the form of a multimodal probability distribution, where
there is one peak at or near the correct redshift, but at least one
other false peak in $(z,T)$-space (see Figs.~\ref{fig:degengals} and
\ref{fig:galspectra}). The astrophysical origin of these multiple
peaks in $(z,T)$-space (and in fact the ability for photometric
redshifts to perform as well as they do) comes from the Lyman
(1216~\AA) and Balmer (4000~\AA) breaks in galaxy spectra. The
transition of these breaks through a set of filters as a function of
redshift provides the change in galaxy color that all photometric
redshift techniques key into. If our photometric observations are
sufficient to measure the galaxy flux for both breaks then the
templates usually have the power to determine the galaxy redshift
unambiguously, and there is no degeneracy.  This is often not the case
for optical surveys. For example, the GOODS survey filters cover the
range of approx.\ 3700--10000~\AA, so the Lyman break is only
observable for galaxies with $3 < z < 6$, and the 4000~\AA\ break
should be observable for galaxies with $z < 1.5$. With only sufficient
spectral range to identify a break in the spectrum, but not to classify
which break it corresponds to, the template fitting procedure will
produce multiple peaks in the $P(z,T)$ distribution. The VVDS adds the
infrared filters ($J$ and $K_s$ bands) which should help identify the
4000~\AA\ break out to higher redshifts and, therefore, disambiguate
the spectral breaks; however, at the depth of the $I$-band selected
and $I<24$ magnitude limited sample the infrared observations do not
have sufficient signal-to-noise to constrain the infrared colors. In
the surveys under study, there are almost no non-QSO/AGN galaxies
above $z = 2$ that have spectra which meet our quality criterion, so
most of the information available to our template fits comes from the
4000~\AA\ break.

In the following subsections we examine how two different methods can
break this degeneracy: first, our approach using an
empirically-measured SB($z$) relation, and then a widely-used method
using the apparent magnitude.

\subsection{Surface Brightness Prior}

\subsubsection{Tolman Test}

\label{sec:sbprior}

\begin{figure*}
  \centering
  \includegraphics[width=0.45\linewidth]{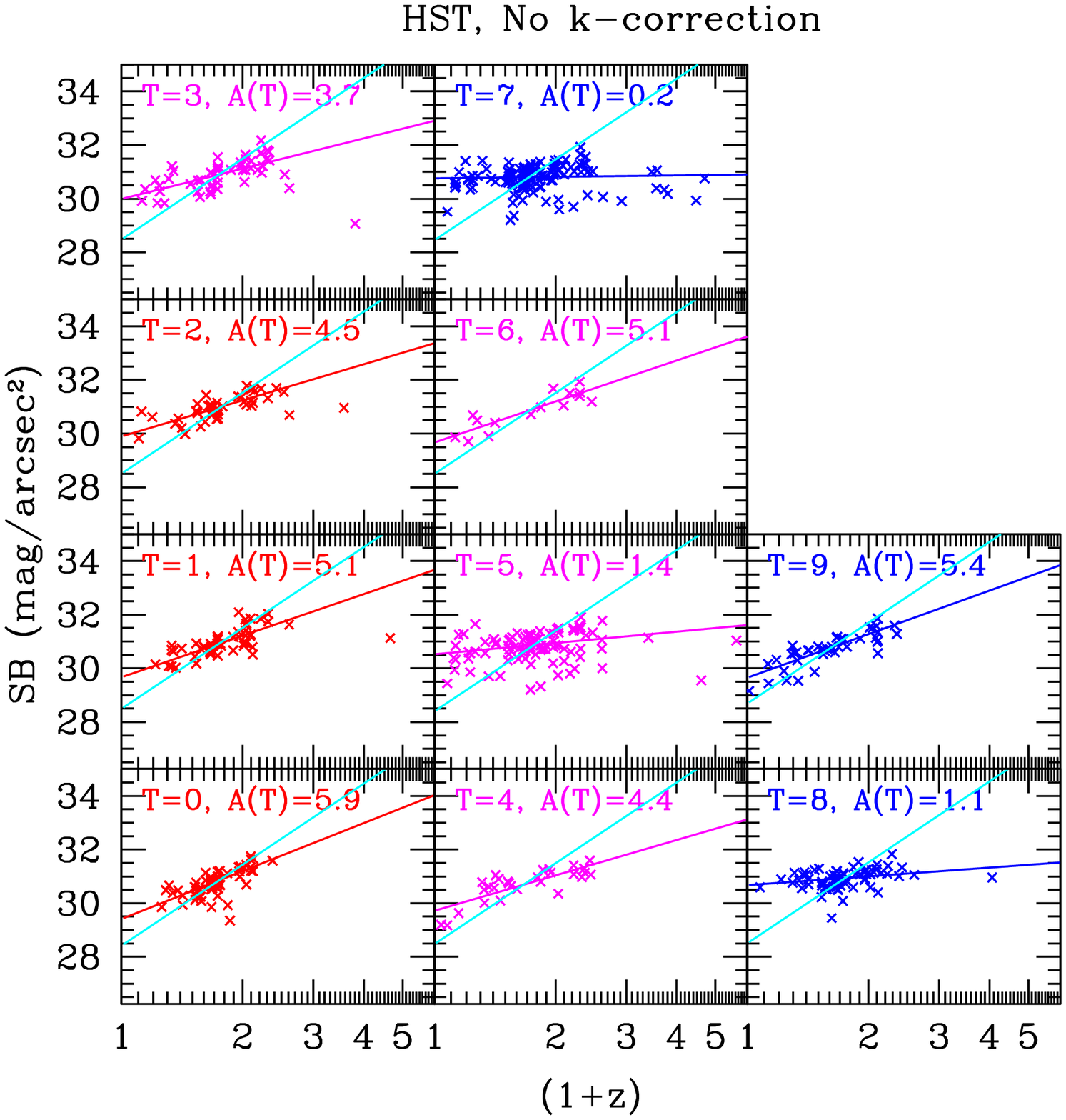}
  \includegraphics[width=0.45\linewidth]{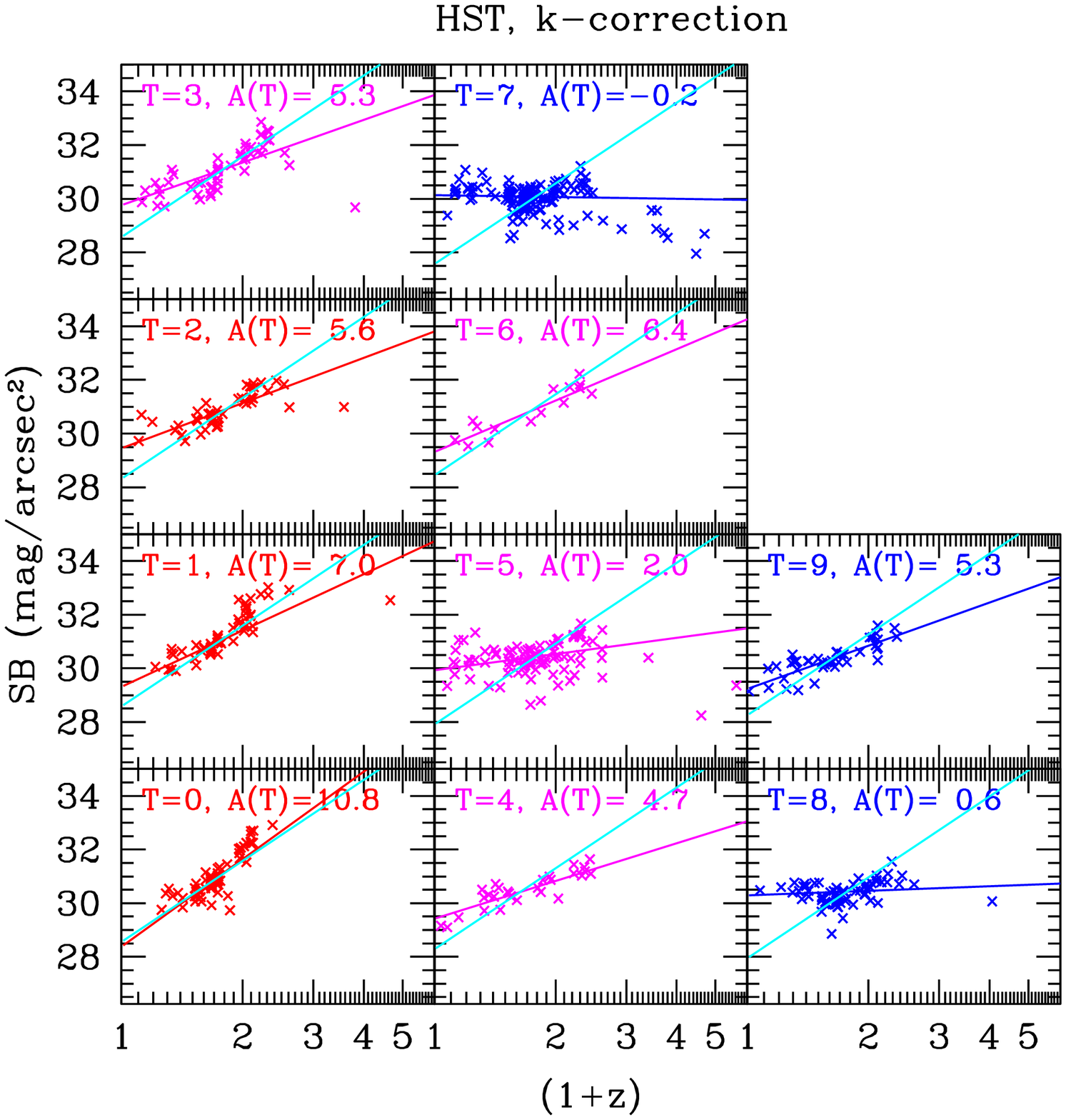}
  \includegraphics[width=0.45\linewidth]{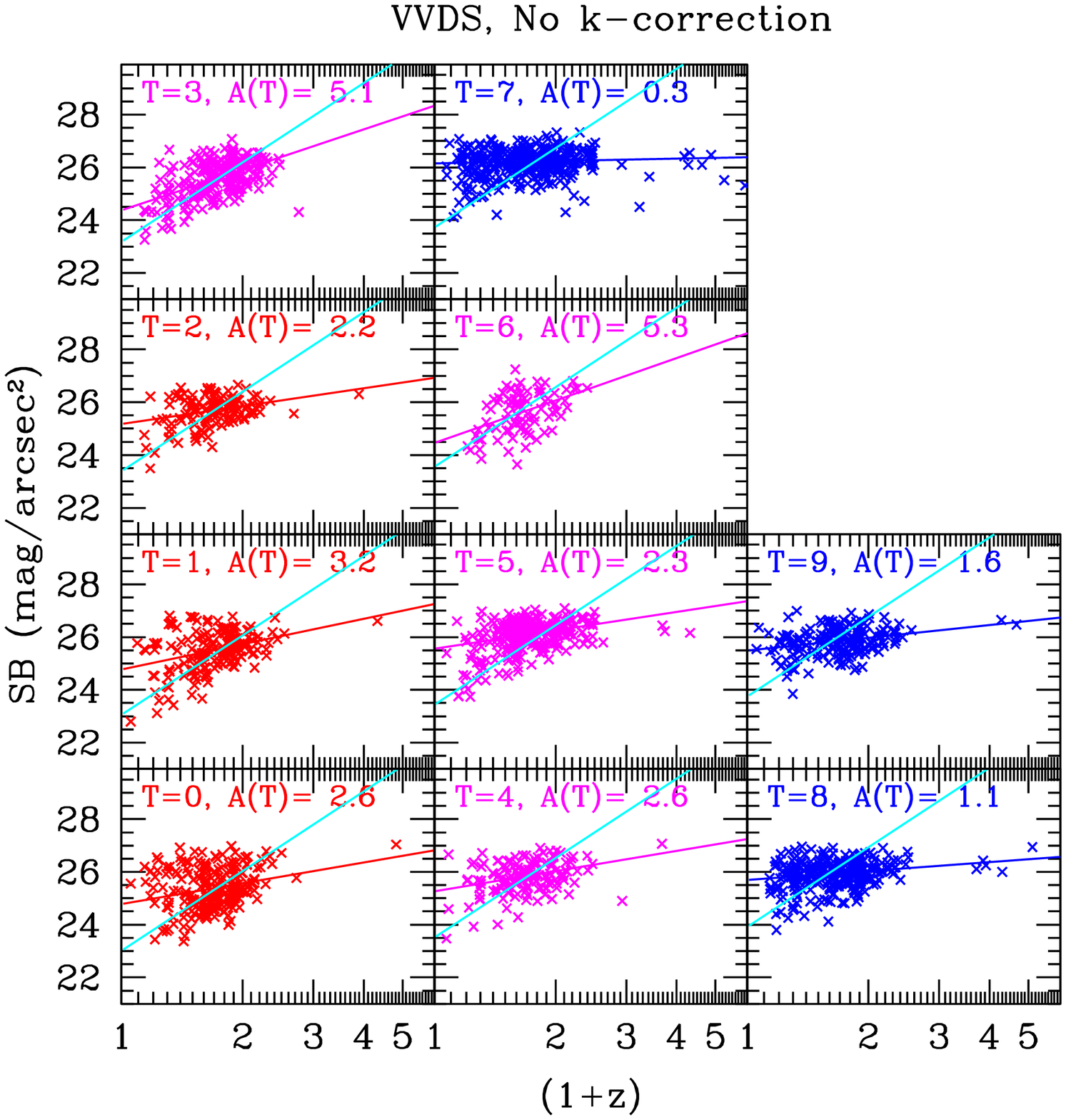}
  \includegraphics[width=0.45\linewidth]{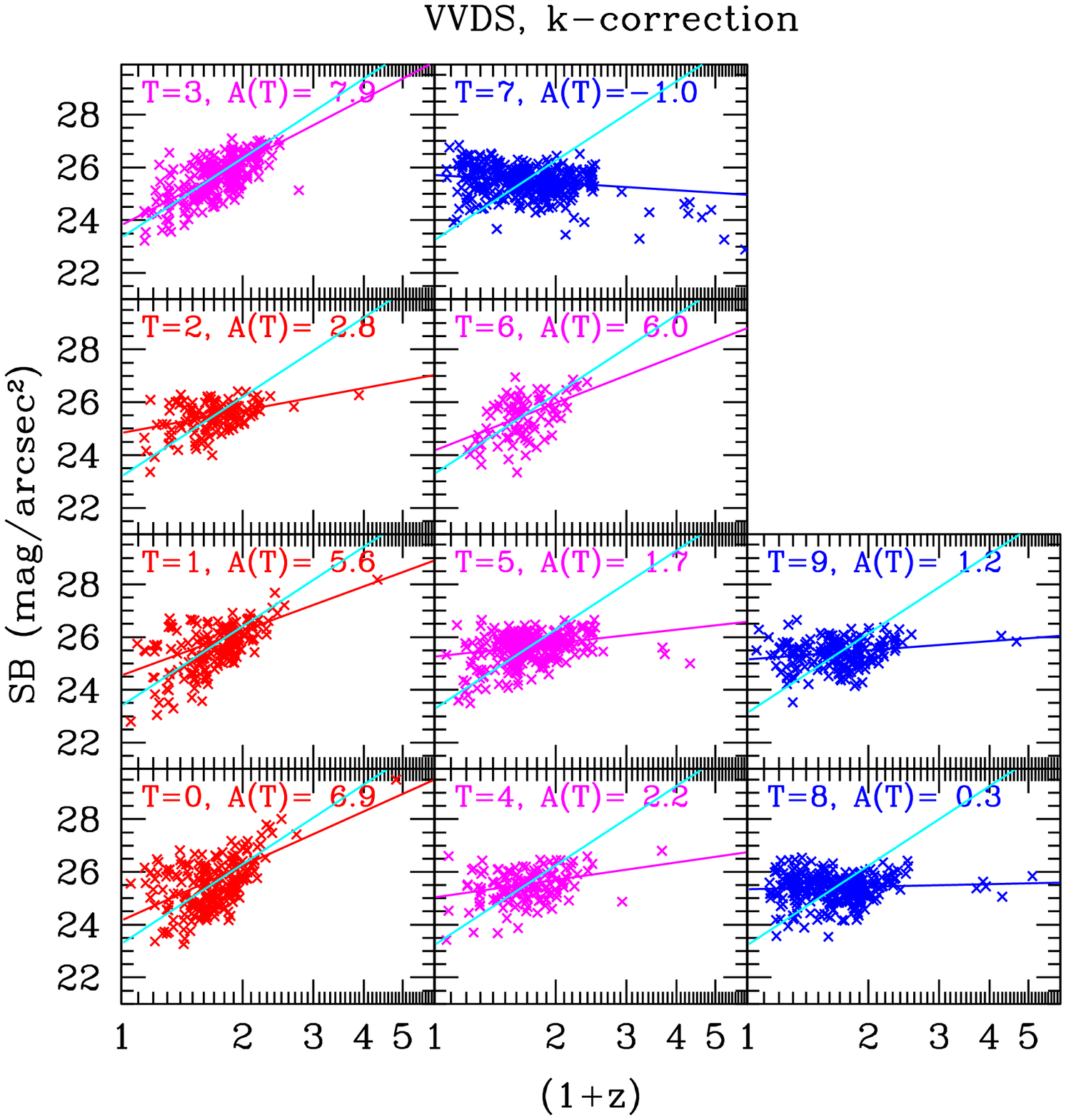}
  \caption{
    Surface brightness (SB) as a function of spectroscopic redshift,
    from space and from the ground, with and without magnitude
    $k$-correction derived from the spectroscopic redshift $z_s$ and
    best-fit template.  The panels are labeled by galaxy type (see
    Fig.~\ref{fig:templates}) and best fit slope of the
    SB-$\log(1+z)$ relation (see Eq.~(\ref{eq:SBfit})). When observed
    from space, the passively-evolving red galaxies are close to the
    $(1+z)^{-4}$ power law predicted by Eq.~(\ref{eq:sbz}), shown by
    the cyan line in each panel.  Because the blue galaxies evolve
    with $z$, they have a shallower slope; this effect becomes more
    dramatic when $k$-corrections are considered.  No attempt has been
    made to compensate for the larger seeing in VVDS.}
  \label{fig:SBz}

\end{figure*}

In this section, we consider the performance of using SB to break the
color-redshift degeneracy.  
The use of SB in constraining photo-z's has been proposed a number of
times, for example, in the ``$\mu$-PhotoZ'' method
\citep{2007AJ....134.1360K}, which is calibrated with the SDSS.  The
authors use one-color photometry and surface brightness to get
photometric redshifts for very red galaxies with $z<0.8$.  While their
method works well the reddest 10--20\% of galaxies, the constraints
worsen for bluer types, and thus it is probably not suitable for a
general photo-z survey.  The SB was also used in \citet{Wray:2007zg}
as part of a larger framework to improve photo-z's for a low redshift
sample of SDSS galaxies.

The surface brightness (SB) of a galaxy is defined (in units of
mag/arcsec$^2$) as:
\begin{equation}
  \label{eq:sb}
  \SB = m+ 2.5\log{A},
\end{equation}
where $m$ is the $I$-band magnitude and $A$ is the angular area.  The
SB increases as the galaxy gets brighter and smaller.  If we let
$\chi(z)$ be the comoving distance to redshift $z$, and recall that $m
\sim 2.5 \log{[\chi^2(z)(1+z)^2]}$ and $\log{A} \sim
\log{[\chi^{-2}(z)(1+z)^2]}$, then neglecting $k$-correction, one can
plug in to Eq.~(\ref{eq:sb}) and derive the evolution of the SB with
redshift:
\begin{equation}
  \label{eq:sbz}
  \SB(z) = 10\log{(1+z)} + {\rm const.}
\end{equation}
Equivalently, in flux units we get $\SB(z)\propto
(1+z)^{-4}$.  This is how we would expect the SB to
behave in an expanding universe for passively evolving galaxies for
which there is no magnitude or size evolution with $z$.  An
interesting point is that $\SB(z)$ has no explicit dependence on the
cosmology; in particular it is independent of the luminosity or
angular diameter distances, unlike the magnitude-redshift relation.
In fact, the SB($z$) relation has been used to test the hypothesis
that the universe is expanding, which is known as the ``Tolman test''
\citep{1930PNAS...16..511T, 1934rtc..book.....T, 2001AJ....122.1084L}.
Instead of using the expected surface brightness-redshift relation to
test whether the universe is expanding, we use SB($z$) to constrain
galaxy redshifts.

Just as in \citet{2001AJ....122.1084L}, our galaxy sample does not
universally follow the simple law given in Eq.~(\ref{eq:sbz}), because
populations of galaxies evolve differently with redshift.  The galaxy
template library \citep[see][and
Fig.~\ref{fig:templates}]{1993ApJ...405..538B, Bruzual:2003tq} enables
us to assign spectral types to each galaxy.  We can, therefore,
attempt to derive the magnitude $k$-correction for each galaxy using
its best fit spectral template and estimate of the redshift.
Fig.~\ref{fig:SBz} shows the SB($z$) relation for each galaxy type in
our datasets, using the spectroscopic redshift information for the
$k$-correction.  We can see that, in the panels with $k$-correction,
the redder galaxies are evolving more passively and hence closer to
the $(1+z)^{-4}$ slope, while the bluer galaxies are actively evolving
in such a way that their SB doesn't change substantially with
redshift; from this we infer that there must be significant evolution
effects that cause intermediate and blue galaxies at higher redshift
to be brighter and/or smaller than passive evolution would indicate.
Although they are constrained to lower redshifts ($z<0.8$), we can see
similar behavior in Fig.~3 of \citet{2007AJ....134.1360K}, where the
lower left panel, containing the 10\% reddest galaxies, follows the
expected $\SB \propto (1+z)^{-4}$ relation most closely, with the
strength of the SB($z$) correlation decreasing as bluer deciles
are considered.

In the plots with $k$-corrections, the measured slopes are steeper for
red galaxies; however, we use the non-$k$-corrected SB to determine
$P(z,T|\SB)$.  The reason for this is that the effect of $k$-correction
is to remove the magnitude evolution with $z$, but the templates
already provide this information, so once we have a best-fit template we
don't gain anything from explicitly $k$-correcting; in fact, it
results in a larger scatter in SB($z$), because the uncertainties in
the k-corrections derived from photometric redshifts increase the
scatter in the SB-redshift relation (due to the ``catastrophic
errors''). The $k$-corrected plots in Fig.~\ref{fig:SBz} have a
small scatter because they have used the spectroscopic redshift in
order to determine the $k$-correction, i.e.\ they use extra information
that the photo-z procedure doesn't have.

\begin{figure*}
  \centering
  \includegraphics[width=0.45\linewidth]{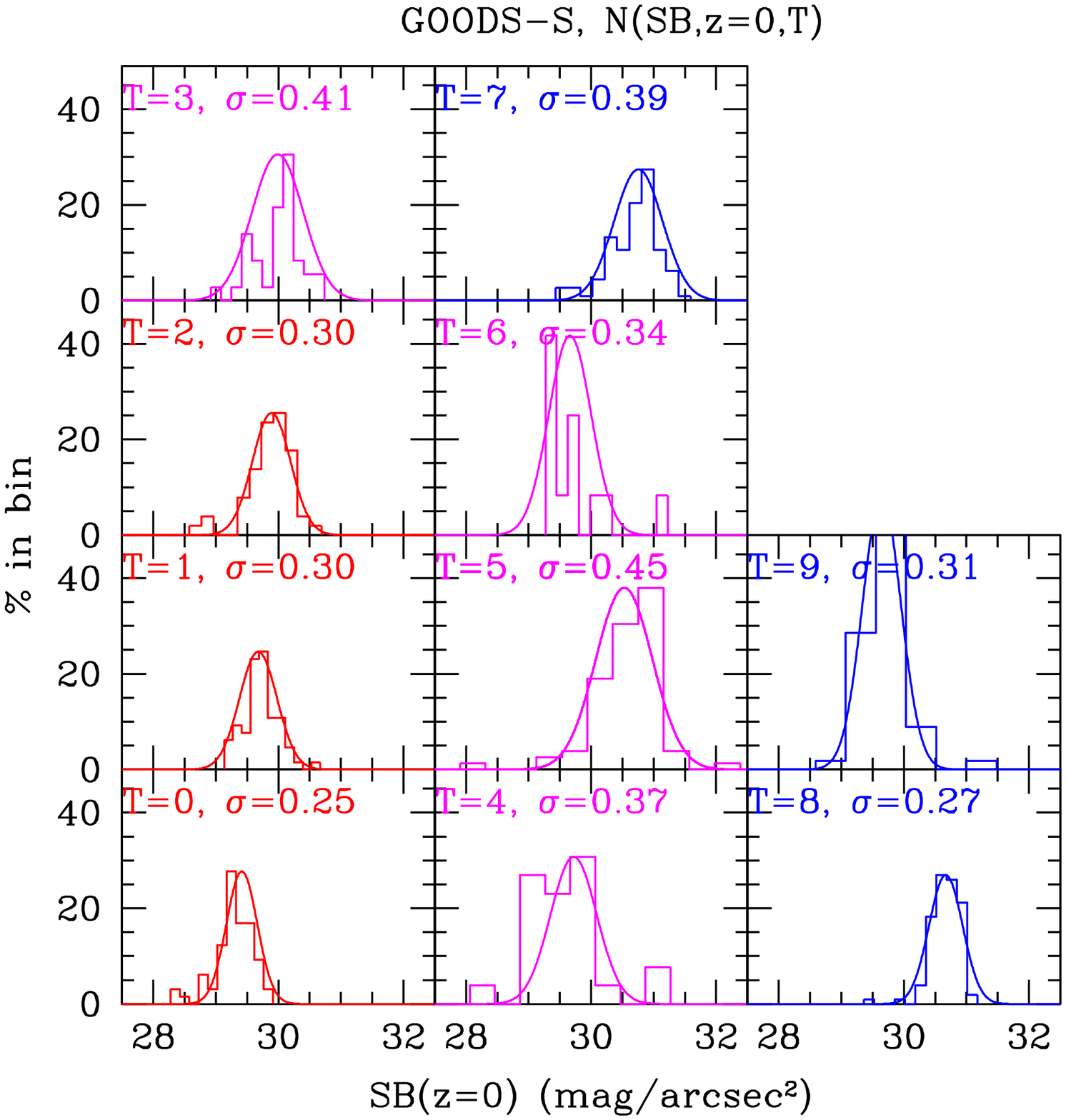}
  \includegraphics[width=0.45\linewidth]{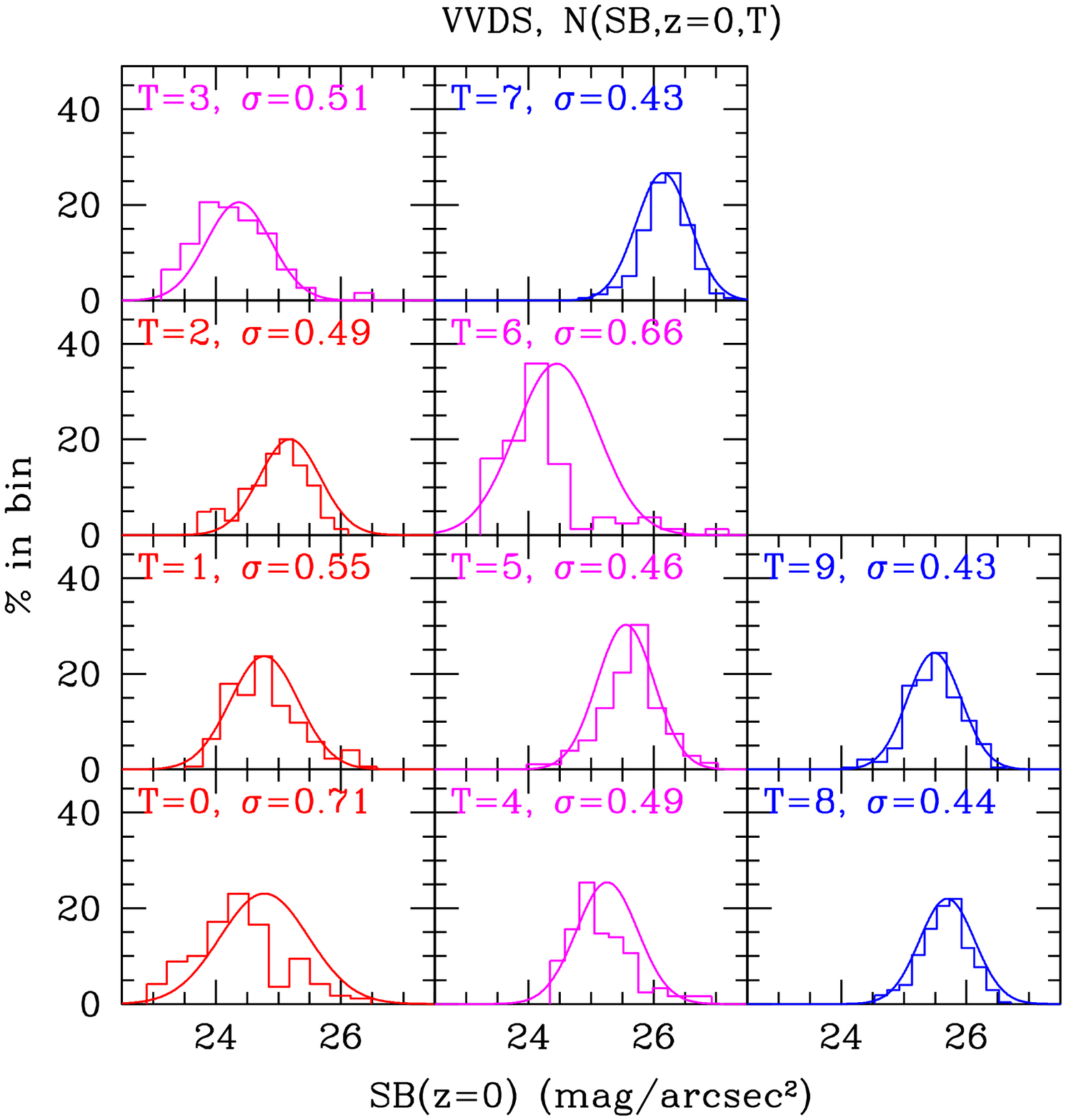}
  \caption{
    $N(\SB,z=0,T)$ with $2\sigma$-clipped Gaussian fits
    for the GOODS-S and VVDS datasets. Using the fit from
    Eq.~\ref{eq:SBfit} we have shifted the galaxy population in each
    template bin to $z=0$ and plotted the resulting distribution.}
  \label{fig:tz0NSB}
\end{figure*}

\begin{figure*}
  \centering
  \includegraphics[width=0.45\linewidth]{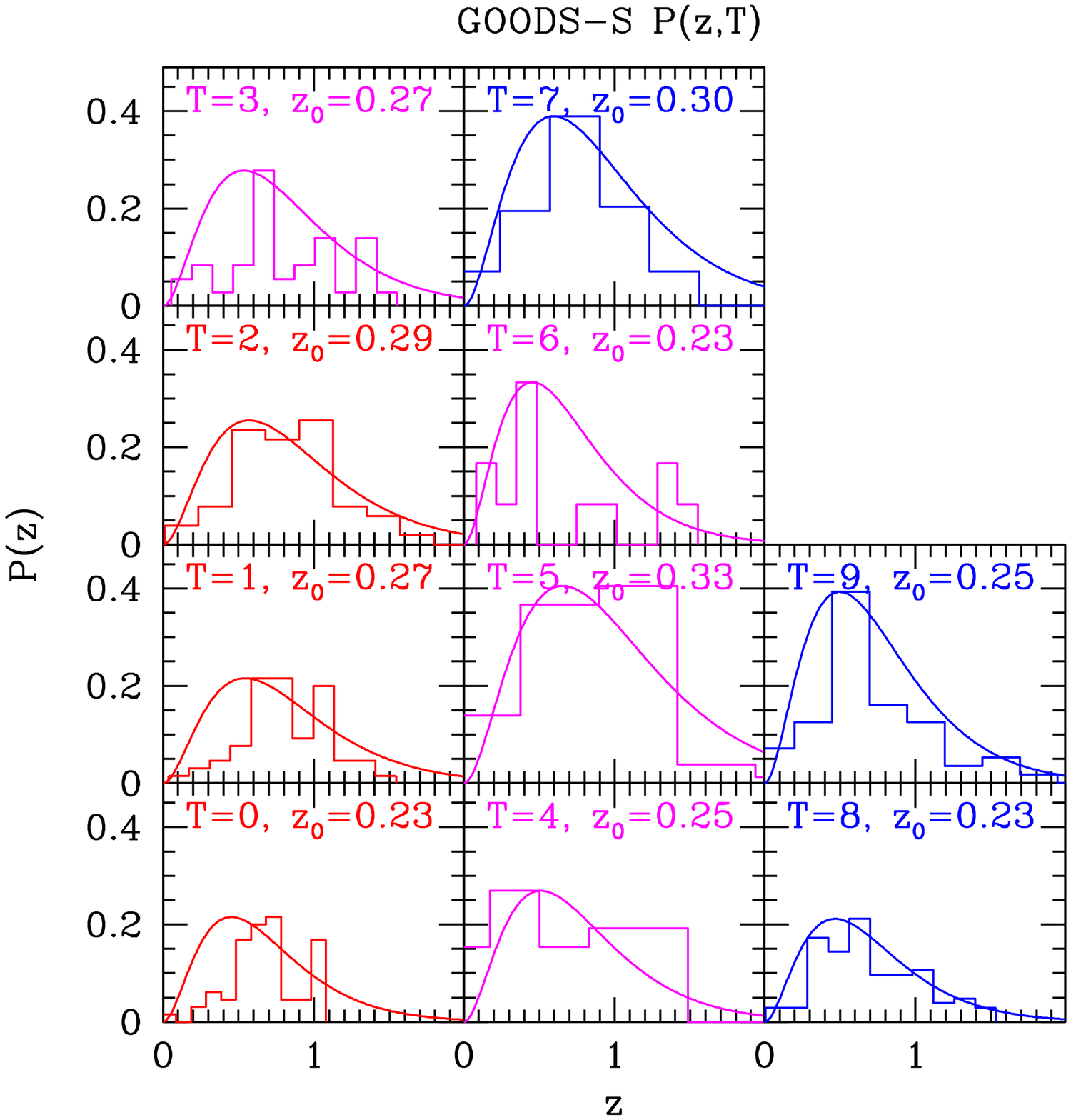}
  \includegraphics[width=0.45\linewidth]{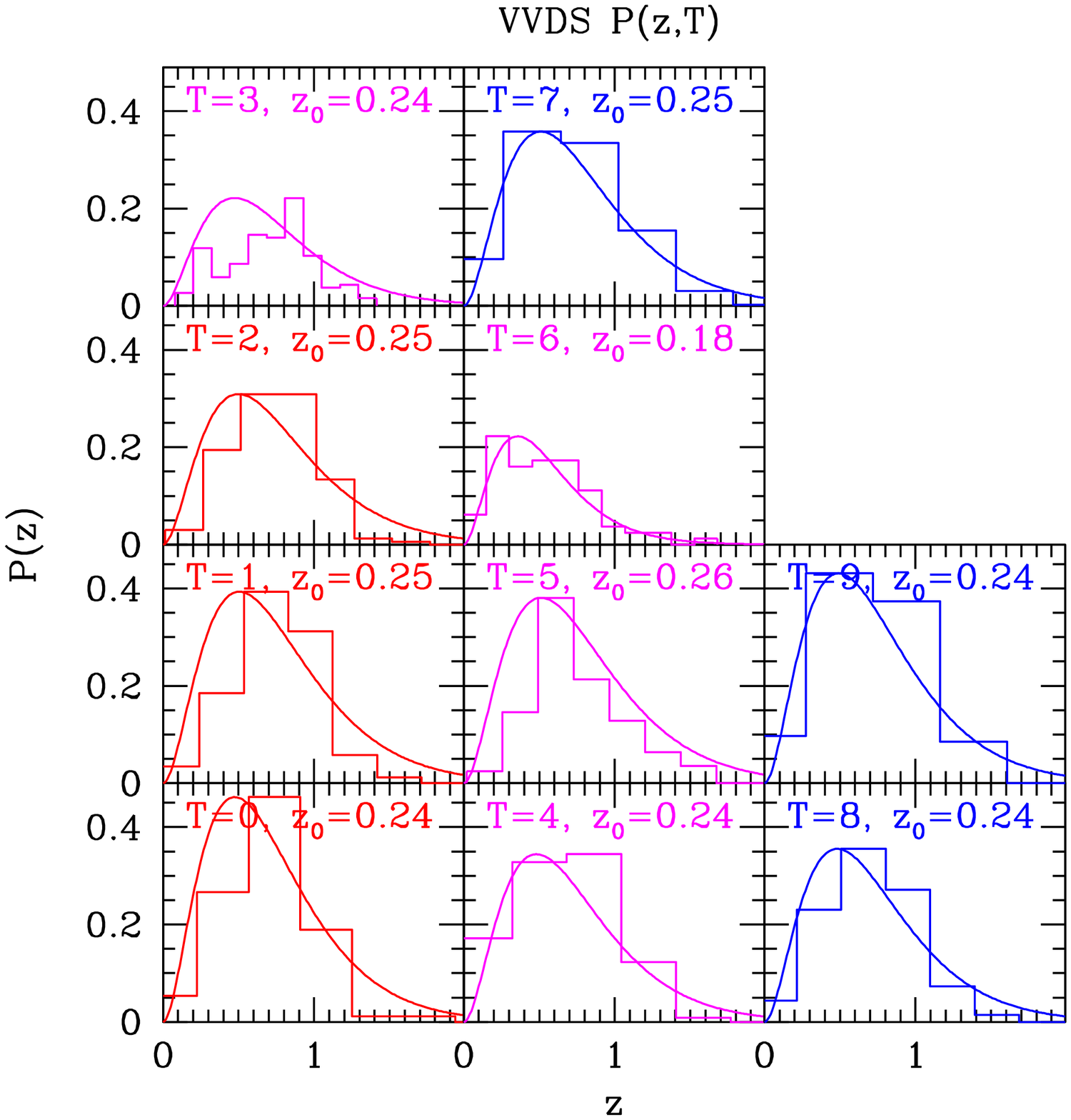}
  \caption{
    Histograms of $P(z,T)$ for GOODS-S and VVDS
    datasets. The solid curves are exponential fits: $P(z,T) =
    (z^2/z_0^3(T)) \exp{[-z/z_0(T)]}$; the parameter $z_0(T)$ is fit
    separately for each type.
  }
  \label{fig:PZT}
\end{figure*}

\subsubsection{Calibration}
We calibrate the SB prior via the following procedure: first, we bin
our galaxies into types $T$; each galaxy in the calibration sample has
a spectroscopic redshift $z_s$ and a measured (non-$k$-corrected)
surface brightness $\SB_\obs$ (computed using Eq.~(\ref{eq:sb}) from
the $I$-band magnitude and size measure).  For each type, we then do a
($2\sigma$-clipped) linear fit
\begin{equation}
  \label{eq:SBfit}
  y(z,T)=A(T)\log{(1+z)}+B(T);
\end{equation}
these fits are plotted in Fig.~\ref{fig:SBz}.  Then, using the fit in
Eq.~(\ref{eq:SBfit}), for every galaxy $i$ we can now compute what its
SB would be if the galaxy were at some redshift $z$ instead of the
observed spectroscopic redshift $z_s$:
\begin{equation}
  \label{eq:SBzT}
  \SB_i(z,T) = \SB_{i,\obs} - 
  A(T) \left[ \log{(1+z_{s,i})} - \log{(1+z)} \right].
\end{equation}
If we then use this equation to move all the galaxies to the same
redshift $z$, we can make a histogram $N(\SB|z,T)$ from the $\SB_i$;
$N(\SB|z=0,T)$, along with Gaussian fits, is shown in
Fig.~\ref{fig:tz0NSB}.  Once we have these fits, we can pick an
arbitrary redshift $z_0$, and using Eq.~(\ref{eq:SBzT}), compute
$\SB_i(z_0,T)$ for every galaxy.  We can further see that
\begin{eqnarray}
  \label{eq:SBzavg}
  \langle \SB(z_0,T) \rangle &=& A(T)\log{(1+z_0)} \\ \nonumber &&+ \langle
  \SB_\obs - A(T)\log{(1+z_s)} \rangle, \\
  \langle \SB(z_0,T)\rangle &=& A(T)\log{(1+z_0)} + B(T),
\end{eqnarray}
where we have used $B(T)$, the maximum likelihood estimator, for $\langle
  \SB_\obs - A(T)\log{(1+z_s)} \rangle$.

The Gaussian fits to the histograms in Fig.~\ref{fig:tz0NSB} give us
the width of the distributions, $\sigma(T)$, for each type.  Now we can
find $P(\SB|z,T)$, which we take to be
\begin{equation}
  \label{eq:PSB_gaussfit1}
  P_i(\SB|z,T) = \frac{1}{\sigma(T)\sqrt{2\pi}} 
  \exp{\left[-\frac{(\SB_\obs-\langle\SB(z,T)\rangle)^2}{2\sigma^2(T)}\right]};
\end{equation}
substituting from Eq.~(\ref{eq:SBzavg}), we get
\begin{eqnarray}
  \label{eq:PSB_gaussfit2}
  P(\SB|z,T) &= &\frac{1}{\sigma(T)\sqrt{2\pi}}\times\\\nonumber
  &&\exp{\left[
      \frac{ (\SB_\obs-A(T)\log{(1+z)}-B(T))^2 } 
      {2\sigma^2(T)} \right]}.
\end{eqnarray}
Finally, we use Bayes' theorem to get the prior distribution
\begin{equation}
  \label{eq:condprob}
  P(z,T|\SB) = P(\SB|z,T)\frac{P(z,T)}{P(\SB)},
\end{equation}
where we have plotted $P(z,T)$ in Fig.~\ref{fig:PZT}, and fit it with
an exponential function
\begin{equation}
  \label{eq:PZT}
  P(z,T)=\frac{z^2}{z_0^3(T)}\exp{\left(-\frac{z}{z_0(T)}\right)}.
\end{equation}
We find $P(\SB)$ for both the GOODS and VVDS samples is well-fit by a
Gaussian distribution.

\subsubsection{Size Measures and Seeing}

\label{sec:seeing}


\begin{table}
  \centering
  \begin{tabular}{c|cc}
    T & Isophotal $\sigma(T)$ & Ellipse $\sigma(T)$ \\\hline
    0 & 0.37 & 0.48 \\
    1 & 0.30 & 0.45 \\
    2 & 0.47 & 0.54 \\
    3 & 0.51 & 0.65 \\
    4 & 0.67 & 0.65 \\
    5 & 0.46 & 0.81 \\
    6 & 0.33 & 1.0 \\
    7 & 0.43 & 0.55 \\
    8 & 0.43 & 0.57 \\
    9 & 0.62 & 0.96
  \end{tabular}
  \caption{
    Effect of size measure on SB measurement in the GOODS data.  For
    each template type, the scatter (in arcseconds) of the SB($z$)
    relations around 
    the best fits are shown for different size 
    measures, on the left using isophotal area, on the right using
    ellipse area.  Using the ellipse instead of isophotal area
    increases the scatter, which is why the precision of the SB
    measurement suffers when using the ellipse area.
  }
  \label{tab:SBz_comp}
\end{table}

\begin{figure*}
  \centering
  \includegraphics[width=0.45\linewidth]{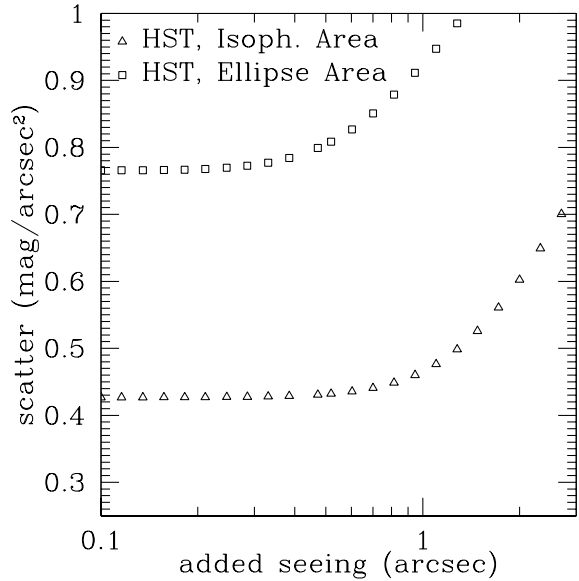}
  \caption{
    Scatter around the $\SB(z)$ relation for red galaxy
    types.  We have plotted the $r_{\rm seeing}$ added (see
    Eq.~(\ref{eq:seeing})) vs the measured $\sigma(T)$ (see
    Eq.~(\ref{eq:PSB_gaussfit1}) and Fig.~\ref{fig:tz0NSB}) for
    $T=[0,1,2]$, using the GOODS data.  The impact of changing size
    measures vs.\ adding seeing is evident: switching from isophotal to
    ellipse area causes the scatter to increase significantly.}
  \label{fig:seeing}
\end{figure*}

\begin{figure*}
  \centering
  \includegraphics[width=0.45\linewidth]{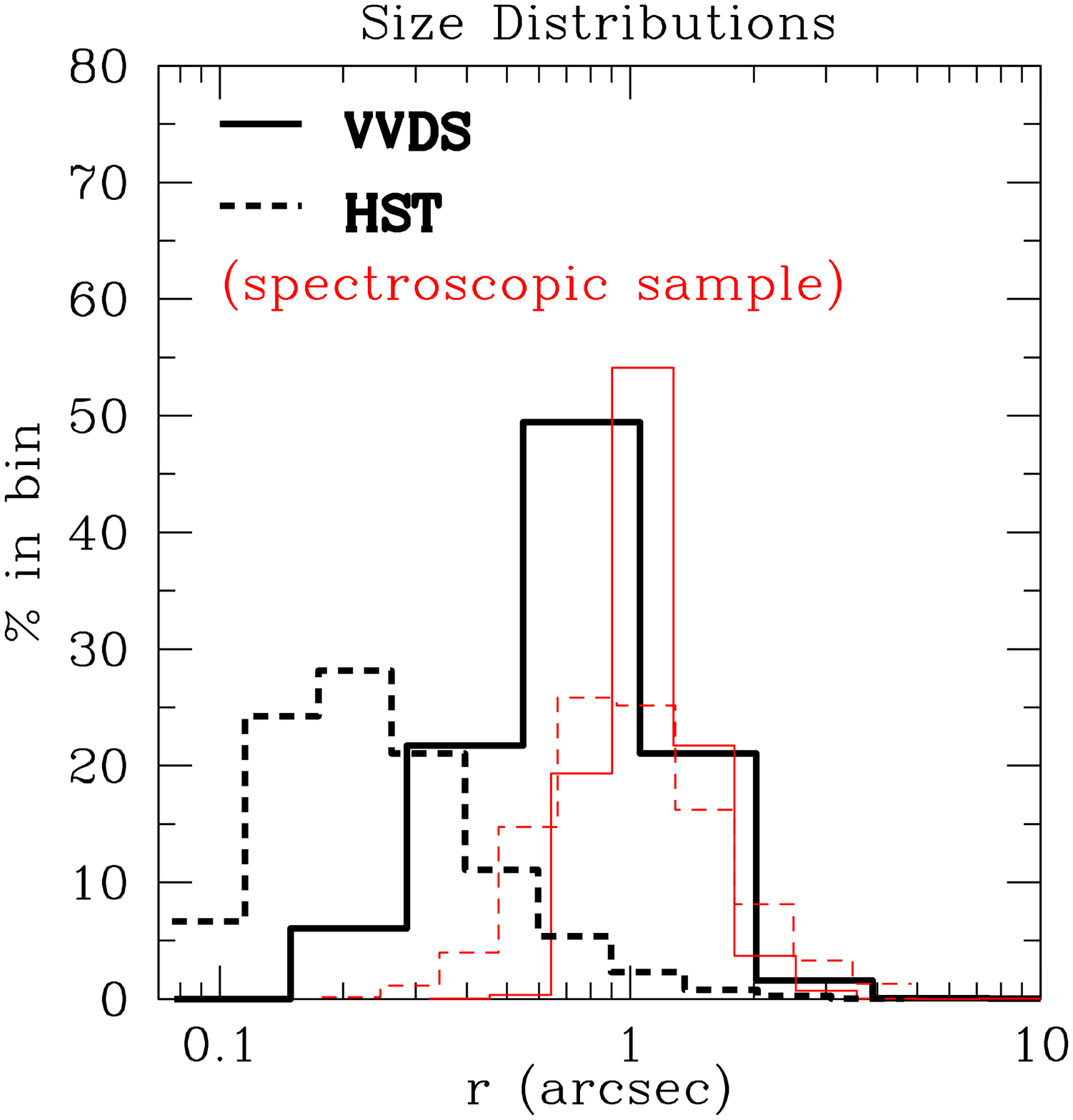}
  \includegraphics[width=0.45\linewidth]{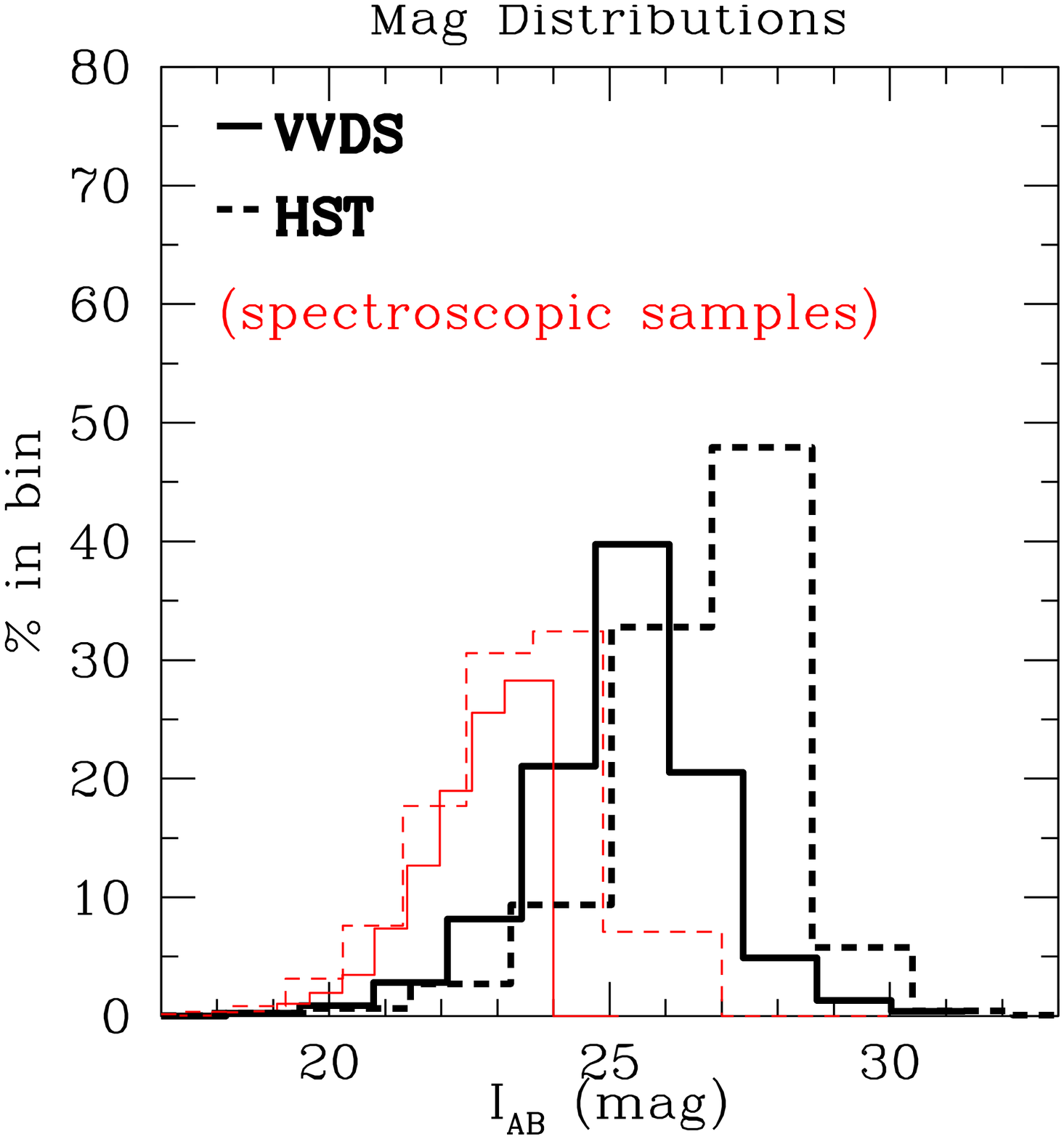}
  \caption{
    {\it Left panel}: size distributions for VVDS
    and GOODS-S galaxies.  The size measure used is $r\equiv\sqrt{\pi
      A B}$, where $A$ and $B$ are the ellipse axes (see text).
    Increased seeing and noise in the ground-based observations shift
    the peak of the size distribution to larger sizes, and also change
    the shape of the distribution.  {\it Right panel}: corresponding
    total magnitude distributions.  The distribution for GOODS-S
    galaxies peaks about two magnitudes fainter than for the VVDS; but
    the spectroscopic samples have very similar magnitude and size
    distributions.
  }
  \label{fig:sizedist}
\end{figure*}

For different public datasets, different size measures are available.
In the GOODS data, the ellipse axes, half-light radius, (filtered)
isophotal area, Gaussian FWHM, and Kron aperture were all cataloged.
For the VVDS data on the other hand, only the ellipse axes were
available.  For the HST data we found that the filtered isophotal
area above the analysis threshold (SExtractor output parameter
\verb=ISOAREAF_IMAGE=) gave the best results for our purposes when
measuring the angular area of the galaxies, giving the tightest
SB($z$) relation.  
In Table~\ref{tab:SBz_comp}, we show that the
scatter of SB($z$) for the space-based data increases by a large
amount for all galaxy types when we use the \verb=A_IMAGE= and
\verb=B_IMAGE= ellipse axis parameters, instead of
\verb=ISOAREAF_IMAGE=, to measure the angular area.

We quantify the impact of the algorithm used to measure area in terms
of the impact of a degradation in the image quality due to seeing. In
Fig.~\ref{fig:seeing} we
apply a transformation to the galaxy sizes, 
\begin{equation}
  \label{eq:seeing}
  r' = \sqrt{r_0^2 + r_{\rm seeing}^2},
\end{equation}
for the red galaxies ($T=[0,1,2]$) and then recompute the RMS
scatter in the $\SB(z)$ relation for those galaxies.  We can see that
in order to make optimal use of SB information for photometric
redshifts, one of the primary issues is that one needs a size measure
that is precise even in the presence of noise and seeing:
Fig.~\ref{fig:seeing} shows that adding 1" of seeing to the isophotal
area measurement increases the scatter in $\SB(z)$ by about 10\%, but
if the ellipse area is used for the measurement, there is about a 80\%
increase in scatter for the same amount of seeing.

We also see in Fig.~\ref{fig:SBz} and Table~\ref{tab:SBz_comp} that the
best-fit slope of the red galaxies when measured from the ground is
much shallower than the best-fit slope when measured from space.
Recall that we expect the blue and intermediate galaxies to have a
shallow slope because they are actively evolving with $z$, but the red
galaxies should have a steep slope even in the presence of seeing.
However, the ground-based observations in Fig.~\ref{fig:SBz} clearly
have a minimum SB cutoff at $\SB\approx 26.5$ mag/arcsec$^2$, below
which no sources are detected.  This means that the increased scatter
due to the seeing and bad size measure naturally leads to a shallower
slope, because larger scatter means that the points in the SB
population which would be measured below the cutoff are instead not
observed, while the points that scatter to larger SB are unaffected.
The net result is similar to that which would be produced by a
Malmquist bias, which will also affect any flux-limited SB($z$)
measurement (our sample should be largely free from the Malmquist
magnitude bias itself because, as illustrated in the right panel of
Fig.~\ref{fig:sizedist}, the spectroscopic sample of galaxies which we
are using is well above the flux limit of the photometric data).

Fig.~\ref{fig:SBz}, Table~\ref{tab:SBz_comp}, and
Fig.~\ref{fig:seeing} together show the performance that we would need
from a ground-based size measurement if we intend it to be useful for
photo-z's.  For the space-based data with the isophotal size measure,
the seeing starts to degrade the scatter around $r_{\rm seeing}\approx
1$", so ideally the seeing from the ground should be less than this in
order to obtain the most constraining power.  Regarding the
sensitivity, we note that purely for the purposes of constraining
photometric redshifts, as long as the calibration sample is
representative of the total sample, the photo-z measurement should be
unbiased, regardless of what the SB detection threshold is.  If for
other purposes we want to measure the slope of SB($z$) in an unbiased
way, i.e.\ with a slope that is not made shallower by the combination
of seeing and noise, a planned survey with a similar depth ($z_{\rm
med}\approx 0.7$) from the ground should be sensitive enough so that
the detection threshold for galaxies is below 26.5 mag/arcsec$^2$;
this requirement will change for a deeper survey, but at the same time
the SB detection threshold naturally becomes more sensitive as
signal-to-noise increases.


\subsection{Magnitude Prior}
\label{sec:magprior}

\begin{figure*}
  \centering
  \includegraphics[width=0.45\linewidth]{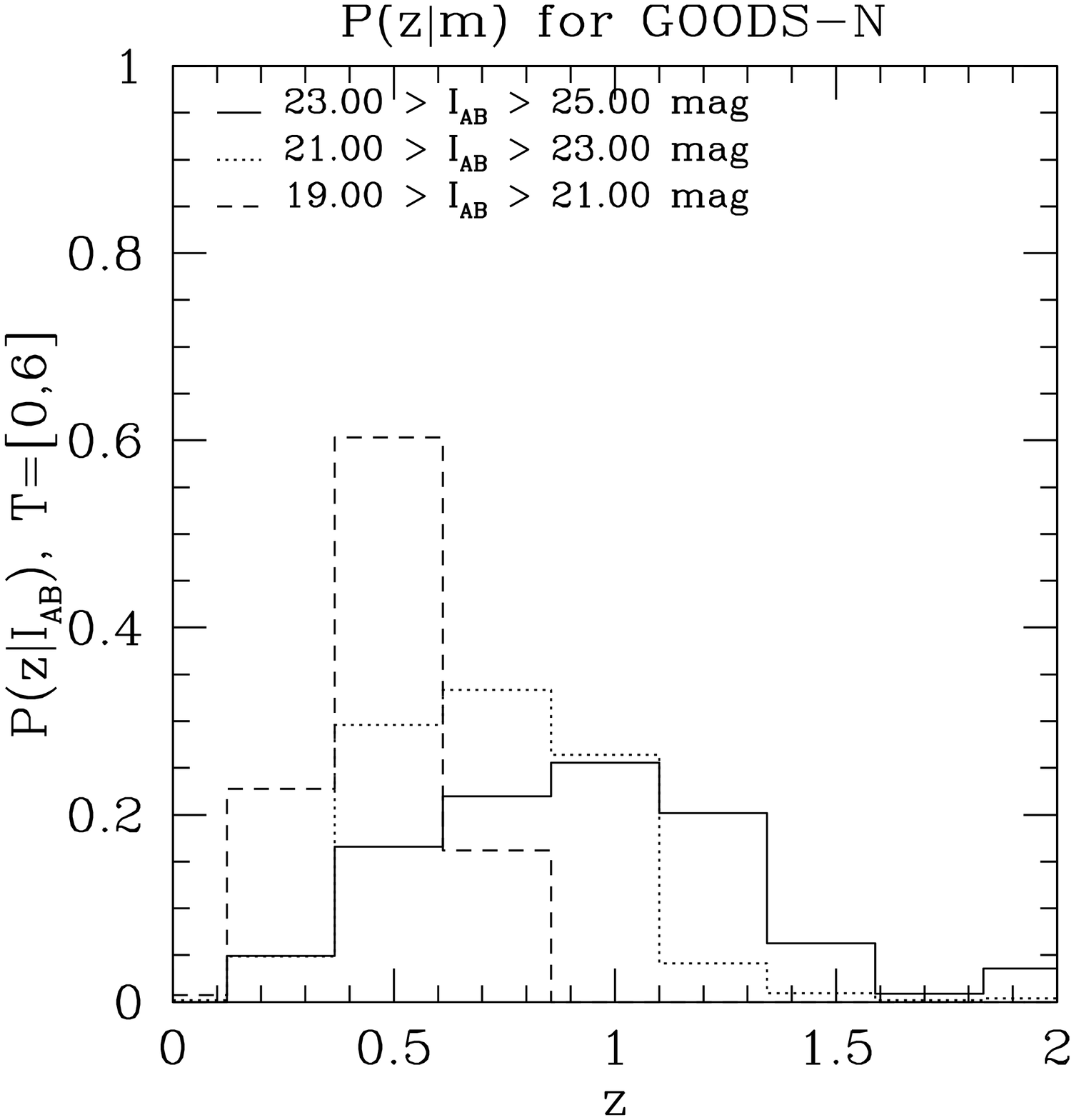}
  \includegraphics[width=0.45\linewidth]{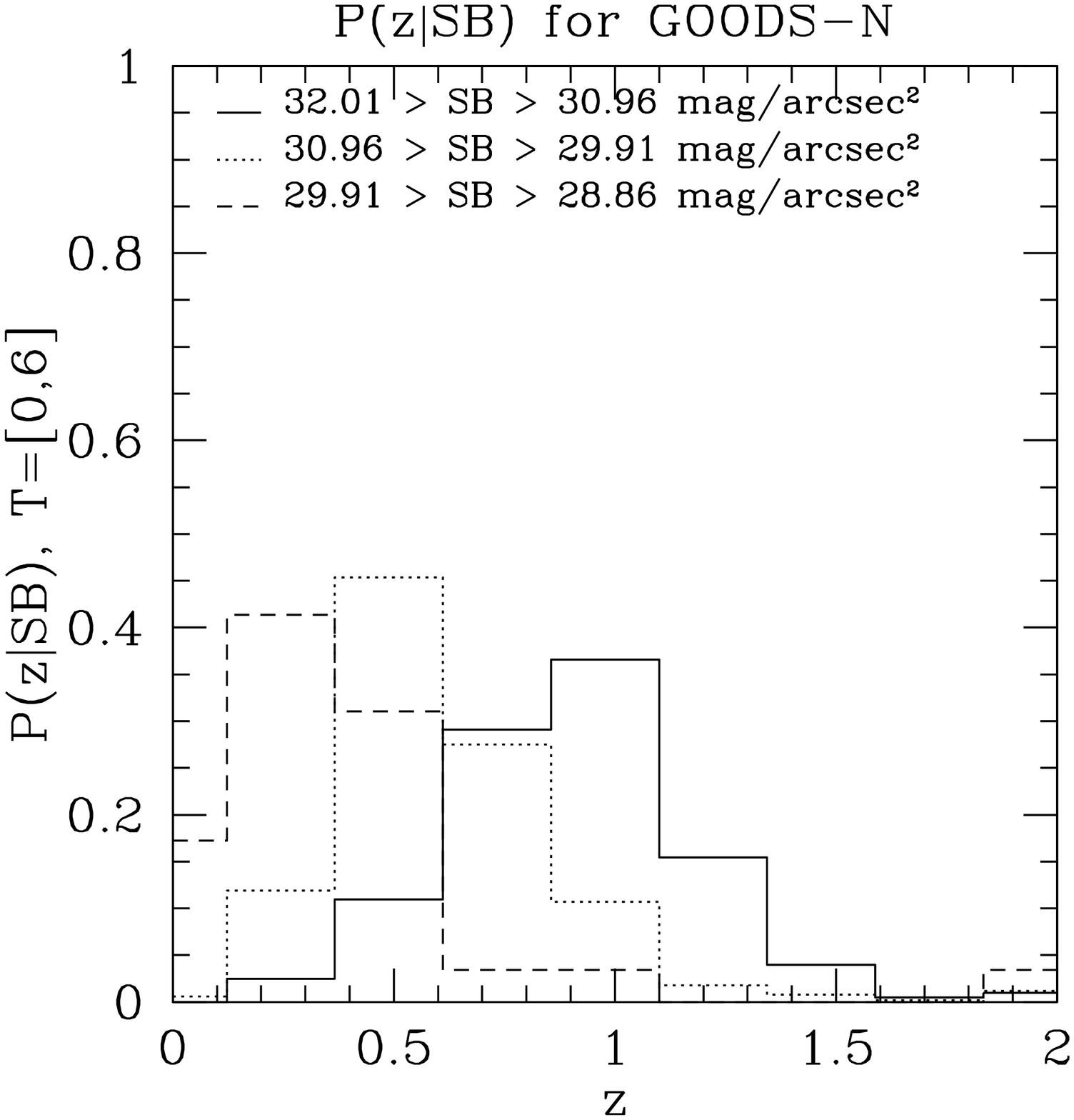}
  \caption{
    {\it Left panel}: 
    $P(z|m)$ for three equal magnitude bins in the GOODS-N data.  As
    the sample becomes fainter, the peak of the distribution moves
    slightly to higher redshift and the histogram becomes appreciably
    broader.  {\it Right panel}: $P(z|\SB)$ for three equal SB bins in
    the GOODS-N data.  The $z$-distribution for $P(z|\SB)$ doesn't
    broaden as much as $P(z|m)$ does for faint samples, and the shift
    of the peak is more pronounced.}
  \label{fig:magprior}
\end{figure*}

Another possible approach to solving the problem illustrated in
Fig.~\ref{fig:galspectra} is to use the overall brightness of a galaxy
as measured in one of the photometric bands; recall that in the
template fitting procedure, in Eq.~(\ref{eq:gal_chisq}), only the
colors and not the brightness of the galaxy are used.  If we take $\Os
\equiv I$ (the $I$-band magnitude) in Eq.~(\ref{eq:pz2}), then we can
use this information to try to break the degeneracies.  Similar
techniques have been used in previous studies \citep[see e.g.][among
others]{Benitez:1998br, Ilbert:2006dp, 2006ApJS..162...20B,
  Feldmann:2006wg, Mobasher:2006fj}.
In \citet{Benitez:1998br} the author demonstrates how to self-consistently
compute a magnitude prior of this type from the galaxy data under
study.  We do not take that approach, and further make the simplifying
assumption that $P(z,T|\Os) = P(z|m)$, i.e.\ galaxy type is
statistically independent of magnitude.  We take our prior information
about the magnitude distribution of galaxies with redshift from the
DEEP2 survey \citep{Coil:2004rj}.  In that survey they found the
following probability density function (PDF) for their spectroscopic
sample:
\begin{equation}
  \label{eq:coilprior}
  P(z|I) \propto dN/dz \propto z^2\exp(-z/z_0),
\end{equation}
where $z_0$ is a function of the $I$-band magnitude.  The parameter
$z_0$ is fitted for in Table~3 of \citep{Coil:2004rj} and we
extrapolate linearly to deeper redshifts from their fit.  From the
shape of $P(z|I)$ distribution plotted in Fig.~\ref{fig:magprior}, we
can anticipate that the effect of applying a magnitude prior will be
to always favor the lower redshift minimum in a $P(z,T)$ probability
distribution.  Since most galaxies really are at low redshift, we will
see that this approach works well; however, using a magnitude prior
exclusively would not be expected to work well if one is interested
primarily in faint galaxies, since a magnitude prior does not provide
much information about faint galaxies, where the prior has a long tail
out to high redshift.  We can see that the SB prior, plotted in the
right panel, does not suffer from this effect, providing a
complementary constraint for faint galaxy samples.  

\section{Results}
\label{sec:results}

In this section we will analyze the results of applying our photo-z
algorithm to each of these datasets with SB priors. We will also
examine the bias and scatter in our measurement errors.

\subsection{Datasets}
\label{sec:datasets}

We use the Great Observatories Origins Deep Survey South (GOODS-S) and
the VIMOS VLT Deep Survey (VVDS) as our calibration samples
\citep{Giavalisco:2003ig, 2005A&A...434...53V, LeFevre:2003mf,
2004A&A...428.1043L, 2005A&A...439..845L}.  The ESO CDF-S master
spectroscopic catalog contains 1115 galaxies with spectroscopic
redshifts, of which we use 603 that have spectroscopic redshift
confidence $\ge 95$\%, i.e.\ we select those that have quality factors
greater than or equal to ``B,'' ``3,'' and ``2.0'' from the following
sources: VLT / FORS2 spectroscopy Version 1.0; VIMOS VLT Deep Survey
(VVDS) Version 1.0; and \citet{2004ApJS..155..271S}.  The VVDS
spectroscopic catalog contains 8981 objects from the VVDS-F02 Deep
field.  Of these, 4180 objects meet our quality criterion ($\ge 95$\%
confidence in the spectroscopic redshift), corresponding to quality
flags 3, 4, 9, 13, 14, and 19.  We use 2090 spectra to calibrate our
method on the ground based data.  We then take the GOODS-N field,
which contains 1814 spectra, and the other half of the VVDS data, to
test our photo-z method on data which are distinct from the
calibration sample.  All results testing the performance of the SB and
combination SB+mag priors are based on these {\it independent}
calibration and test data sets.

The magnitude and size distributions of the sample of galaxies under
study are shown in Fig.~\ref{fig:sizedist}. We can see that while the
spectroscopic samples are very similar, the total size and magnitude
distributions of the galaxies are substantially different for the
ground and space data.  This is because the HST data that we evaluate
have a substantially higher signal-to-noise ratio for the sizes and
magnitudes.  This significantly affects the measurement of SB in each
sample as well as the performance of the SB and magnitude priors.

In order to evaluate the performance of the template-fitting photo-z
algorithm, and any improvements that applying priors produces, we
define as a figure of merit
\begin{equation}
  \label{eq:delta}
  \Delta = \frac{z_p-z_s}{1+z_s},
\end{equation}
which is the fractional error in $1+z_p$, where $z_p$ is the
photometric redshift estimate from Eq.~(\ref{eq:zp}).  For a population of
galaxies, $\langle\Delta\rangle$ and $\sigma_\Delta$ measure the
photo-z bias and the RMS size of the fractional error in
$1+z$.  

Instead of applying a correction to the photometric zero-points and
training the templates using the calibration sample, as in e.g.\ 
\citet[][]{Ilbert:2006dp}, we adopt a simpler and largely equivalent
procedure of fitting the bias (ignoring outliers) on the calibration
set.  First we find a polynomial fit for the bias $F(z_s) = (z_p-z_s)$
for all galaxies with $|\Delta| < 0.5$ in each redshift bin in the
calibration set, and then for every redshift in the test sample we
apply the correction
\begin{equation}
  \label{eq:zpcorrect}
  z_p'=z_p-F(z_p).
\end{equation}
We note that a correction of this type is only expected to work well
in regions adequately sampled by the spectroscopic redshift sample, while a
correction to the magnitude zero-points may be effective over a larger
range.  The black points and histograms in Fig.~\ref{fig:scatter} and
Tab.~\ref{tab:chisq_cuts} show the characteristics of the unweighted
photo-z template fitting algorithm on the test sample after the
correction has been applied.

\subsection{Performance without Priors}
\label{sec:unweighted}

\begin{figure*}
  \centering
  \includegraphics[width=0.45\linewidth]{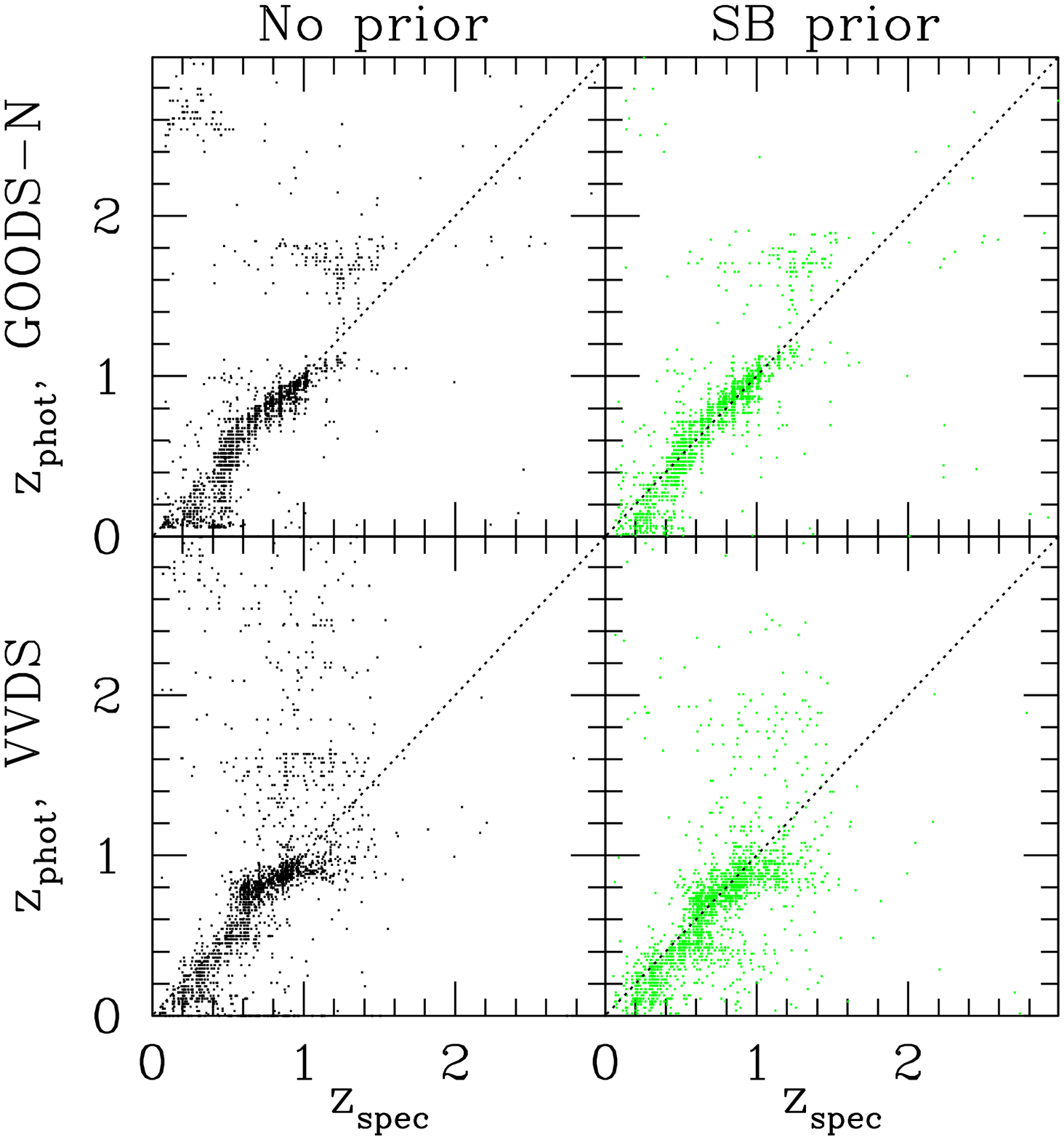}
  \includegraphics[width=0.45\linewidth]{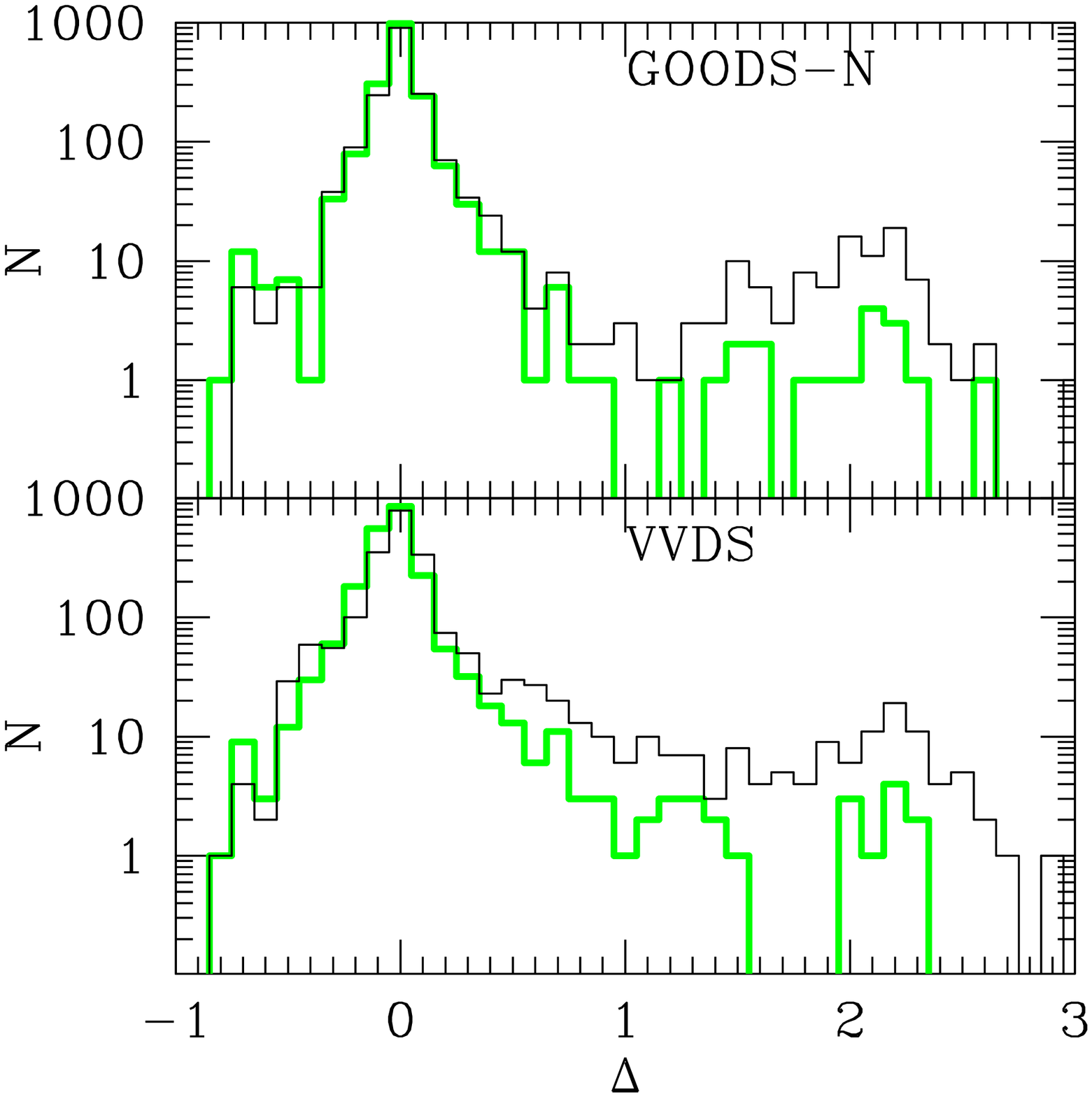}
  \caption{
    {\it Left panel}: Scatter plot of spectroscopic vs photometric
    redshift.  {\it Right panel}: corresponding histogram of
    fractional error in $(1+z)$, $N(\Delta)$, where $\Delta \equiv
    (z_p-z_s)/(1+z_s)$.  The green points and thicker histogram have
    the SB prior, while the black points/histogram have no prior
    applied.  The lower panels contain the ground based data, 2091
    galaxies from the VVDS survey, while the upper has 1814 HST
    galaxies from the GOODS-N field.  In each case these data are
    separate from the calibration datasets.}
  \label{fig:scatter}
\end{figure*}

The performance of the template-based photometric redshift estimator
without priors on the GOODS and VVDS data sets is comparable with that
found by other groups \citep{Mobasher:2006fj,Ilbert:2006dp}. The
dispersion in the photometric redshift relation, defined at the
68$^{\rm th}$ percentile of the distribution, has a scatter of
$\sigma_{68}\le 0.08$ for redshifts $z<1.3$, and $\sigma_{68}\le 0.14$
at redshifts $z>1.3$.  If all of the data, including outliers, is
considered, the dispersion of the photo-z relation is substantially
larger ($>0.45$), which demonstrates the non-Gaussian nature of the
photometric redshift errors, which arises due to the presence of the
outliers or degenerate points (see Fig.~\ref{fig:scatter}).  As noted
by \citet{Mandelbaum:2007dp}, the small bias in the uncorrected (i.e.\ 
no correction of the form Eq.~(\ref{eq:zpcorrect})) photo-z's without
priors is due to a fortuitous cancellation: the outliers which scatter
to high $z$ cancel the bulk of the distribution which is biased low.

As discussed in Sec.~\ref{sec:priors}, the dominant failure mode for
this kind of photo-z procedure is mistaking a 4000~\AA\ break for a
Lyman break, which is visible on the scatter plots in
Fig.~\ref{fig:scatter} as a region in which $z_s \lesssim 1$ but $z_p
\gtrsim 2$; this outlier population is composed of galaxies whose
probability distribution $P(z,T)$ has multiple peaks (e.g.\ the
galaxies in Fig.~\ref{fig:degengals}).  Another type of failure can
happen when points on both sides of the rest-frame 4000~\AA\ break are
not observed in the survey filters.  We can see this in the VVDS
where the performance of the unweighted algorithm suffers at redshifts
at $z\gtrsim 1.2$ due to the lack of $z$-band data: at these redshifts, the
VVDS outlier fraction and $\sigma_{68}$ increase sharply as the
4000~\AA\ break redshifts out of the $I$-band; unfortunately, the $J$
and $K_s$ bands are much shallower than their optical counterparts and
so do not help constrain high-$z$ galaxy photo-z's.  A similar
phenomenon is visible in both VVDS and GOODS for galaxies with
$z<0.4$, due to the lack of deep $u$-band imaging.


\subsection{SB Prior Performance}
\label{sec:prior_performance}

\begin{figure*}
  \centering
  \includegraphics[width=0.45\linewidth]{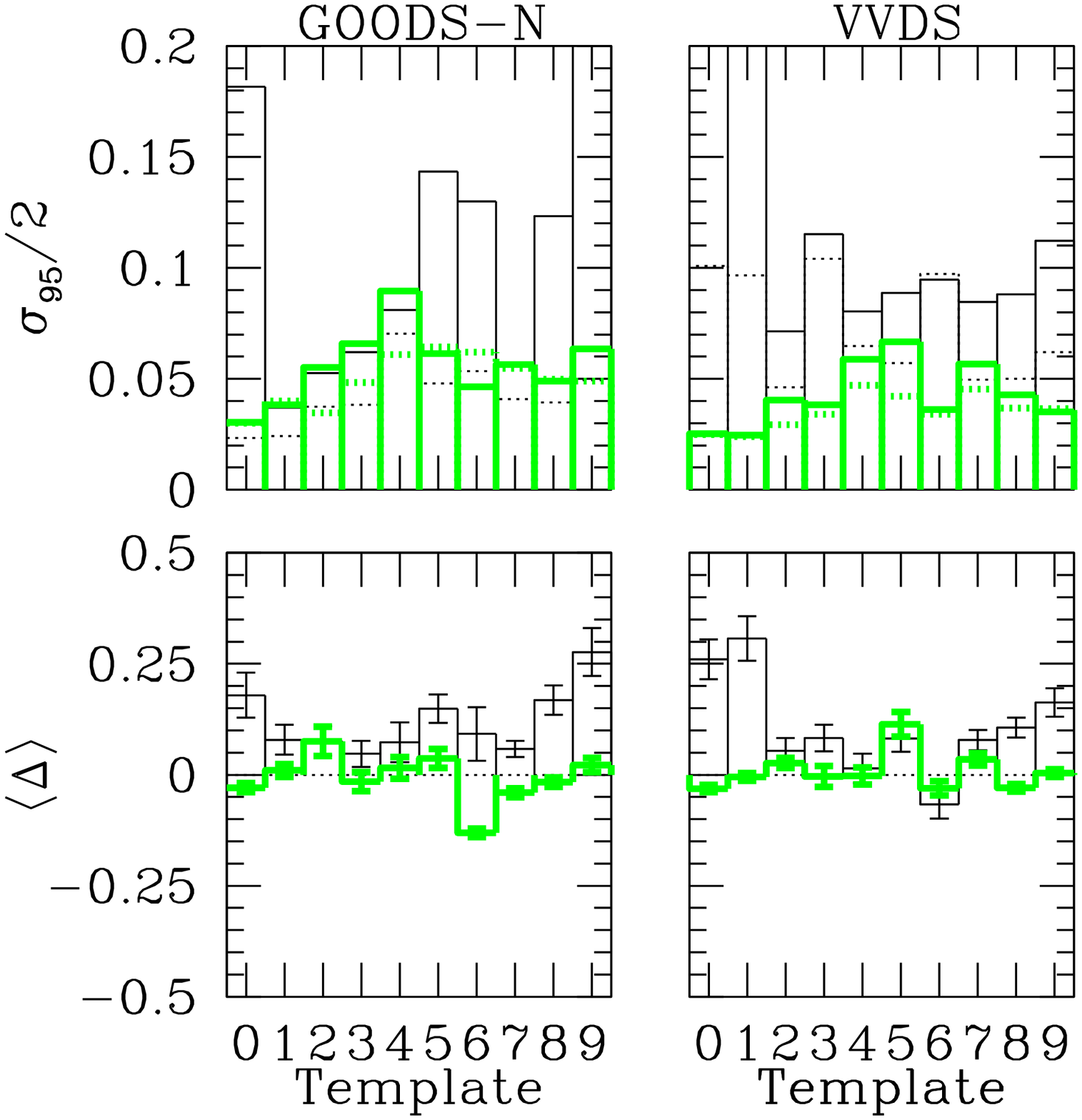}
  \includegraphics[width=0.45\linewidth]{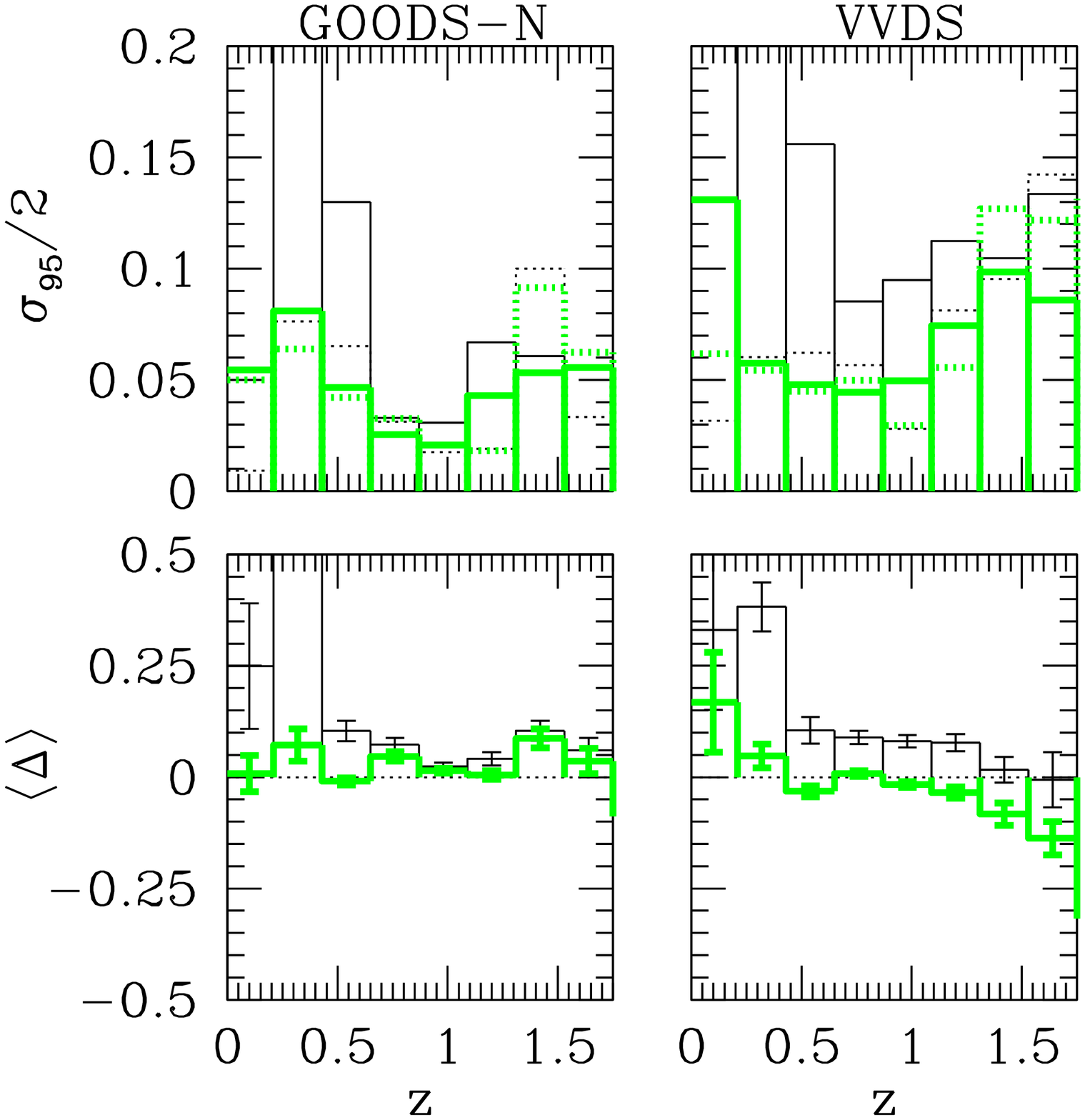}
  \caption{
    {\it Left panel}: bias ($\langle\Delta\rangle$) and scatter at the
    (solid) 95$^{\rm th}$ or (dotted) 68$^{\rm th}$ percentile as a
    function of template.  {\it Right panel}: bias and scatter as a
    function of redshift.  The thin black histograms have no prior
    applied, while the thick green histograms show the results with
    the SB prior.  The SB prior improves both scatter and bias for
    essentially all templates and redshifts.  The improvement at high
    redshift is greater for the GOODS data.
  }
  \label{fig:bias_byzt}
\end{figure*}

Fig.~\ref{fig:bias_byzt} shows a breakdown of the impact of the SB
prior on the photo-z error distribution in different redshift and type
bins; we can see that the SB prior improves bias and scatter on the
independent test sample in almost all redshift and template bins.  We
characterize the scatter in the photometric redshifts as the
dispersion given at the 95$^{\rm th}$ and 68$^{\rm th}$ percentiles of
the distribution, which are denoted by solid ($\sigma_{95}/2$) and
dotted ($\sigma_{68}$) lines respectively; the two measures of scatter
differ in how sensitive they are to outliers.  After the SB prior is
applied, we obtain $\sigma_{95}/2 < 5$\% for the GOODS-N and
$\sigma_{95}/2 < 8$\% for the VVDS data in each bin in the redshift
range $0.4<z<1.3$. The largest improvement in $\sigma_{95}$ is in the
low redshift bins $z<0.6$ for the GOODS and $z<1.3$ for the VVDS data,
because the SB prior eliminates the largest number of outliers in
these bins; we can see that outliers increase the scatter at $z<0.6$
by a factor of 2--3 for the GOODS data and by a factor of 1.5--2 for
redshifts $z<1.3$ in the VVDS data. At higher redshifts the photo-z
algorithm performs better on the space-based data from GOODS, due to
the higher signal-to-noise, better size measurement, and $z$ band
data. For the templates, the largest improvements in accuracy and
precision are gained for the reddest (types 0 and 1) and bluest (types
8 and 9) galaxies.  For the very old and red galaxy types, this
  is to be expected, because they evolve passively with redshift.  As
  we see in Fig.~\ref{fig:SBz}, this means that galaxy types that have
  steep slopes close to $10 \log{(1+z)}$ provide the strongest
  redshift constraints.  For the very young and blue galaxies, we see
  from Fig.~\ref{fig:tz0NSB} that the absolute SB distribution is very
  tight for these types, so even though they have a shallow SB($z$)
  slope the prior is able to provide a constraint on their redshift.
  
  We can gain some intuition about the physical reasons for the
  difference between the SB($z$) relation for the red and blue galaxy
  types by considering the surface brightness of a galaxy population
  with characteristic luminosity $L_0$ and size $R_0$:
  \begin{eqnarray}
    \label{eq:SBevloution}
    \SB 
    &\propto& \frac{L_0}{R_0^2} \frac{(1+z)^q}{(1+z)^{2\rho}} \frac{1}{(1+z)^4},
  \end{eqnarray}
  where the exponents $q$ and $\rho$ describe the luminosity and size
  evolution of the sample, respectively.  Studies of starbursting
  galaxies \citep{Dahlen:2006sy} have indicated that the size
  evolution of these galaxies is fit by $\rho\approx -1$.
  \citet{Willmer:2005fk} find that $M^*_B$ for blue galaxies in the
  DEEP2 survey brightens by approximately 1.3 per unit redshift, i.e.\ 
  $q\approx 0.5$.  Therefore, ignoring selection effects, we see that
  for blue galaxies a combination of size and luminosity evolution
  should partially cancel the geometric factor of $(1+z)^{-4}$.  We
  observe this in our data: the general trend in Fig.~\ref{fig:SBz} is
  that for the bluer galaxies, evolution and selection together
  produce an effect that cancels the $(1+z)^{-4}$ evolution, while for
  the red galaxies, the effect is much less pronounced.  This is
  consistent with our knowledge that red galaxies evolve much less
  than blue galaxies do.  

We can also see how the bias $\langle\Delta\rangle$ is improved in
each redshift and template bin.  By eliminating the high-$z$ outliers,
we achieve bias of less than 5\% in each redshift bin in the range
$0.4<z<1.3$ for the GOODS data, and less than 4\% for the VVDS in the
same range.  When the SB prior is applied, the photo-z estimator
continues to perform well outside this redshift range on the GOODS
sample, while for the VVDS, there is larger bias at very low and very
high $z$.  This is due primarily to the fundamental limitations of the
VVDS survey bands (no deep infrared data) we have discussed
previously, as well as the issues that can affect the SB prior in
ground-based observations: seeing, sensitivity, and size measurement
(see Sec.~\ref{sec:seeing}).

Table~\ref{tab:chisq_cuts} shows the global bias ($\langle \Delta
\rangle$), scatter ($\sigma_\Delta$, $\sigma_{95}$, $\sigma_{68}$),
and fraction of outliers ($N(|\Delta| > 0.2)$) for the SB, magnitude,
and combination priors for our GOODS and VVDS test samples.  Both the
magnitude and SB priors have comparable effects on the fraction of
outliers in the photo-z relation, reducing it by a factor of 2 before
any further cuts. If the goal of an analysis is as small a scatter as
possible then applying both the magnitude and SB prior produces the
best results for both GOODS and VVDS.

We also show in Table~\ref{tab:chisq_cuts} a simple way to reduce the
outlier fraction in the results of our photo-z algorithm.  First we
sort the galaxies on the value of the posterior probability
$P(z,T|\Cs,\Os)$, i.e.\ the final $\chi^2$, at the best-fit $(z,T)$.  Then
the subsequent rows after the first in each subsection of
Table~\ref{tab:chisq_cuts} are produced by cutting a certain fraction
of the worst-fitting (highest $\chi^2$) galaxies from the sample and
then evaluating the bias and scatter of the remainder.  For the GOODS
data, this produces an improvement in the scatter and outlier fraction
as well as the bias; however for the VVDS, making this cut entails a
tradeoff between bias and scatter.  In both cases we can achieve an
outlier fraction of $\le 2$\% if we are willing to sacrifice 20\% of
the sample.

\begin{table*}
  \centering
  \parbox{0.45\linewidth}{\centering {\large VVDS} \\
  \begin{tabular}{rc|c@{ $\pm$ }l|ccc|c}
    Prior & 
    cut  
    & $\langle\Delta\rangle$ &
    $\frac{\sigma_\Delta}{\sqrt{N}}$  
    & $\sigma_\Delta$ & $\sigma_{68}$ & $\sigma_{95}/2$ & outl. \\\hline
No prior&   100  &   11.9  &   1.1  &   48.9  &    6.2  &   11.7  &   15.7    \\ 
&    99  &   11.8  &   1.1  &   48.7  &    6.2  &   11.6  &   15.7    \\ 
&    95  &   12.0  &   1.1  &   48.8  &    6.3  &   11.8  &   16.1    \\ 
&    80  &   13.5  &   1.2  &   50.0  &    6.6  &   12.7  &   17.8    \\\hline 
Mag&   100  &   -0.4  &   0.5  &   25.0  &    5.4  &    5.8  &    7.1    \\ 
&    99  &   -1.4  &   0.4  &   20.1  &    5.4  &    5.6  &    6.6    \\ 
&    95  &   -1.7  &   0.4  &   17.2  &    5.1  &    5.0  &    5.7    \\ 
&    80  &   -3.3  &   0.2  &   10.0  &    4.4  &    3.8  &    1.5    \\\hline 
SB&   100  &   -1.4  &   0.5  &   24.0  &    5.1  &    5.7  &    6.2    \\ 
&    99  &   -2.2  &   0.4  &   19.8  &    5.1  &    5.5  &    5.8    \\ 
&    95  &   -2.6  &   0.4  &   17.4  &    4.9  &    5.0  &    4.7    \\ 
&    80  &   -3.5  &   0.3  &   11.1  &    4.3  &    4.0  &    1.7    \\\hline 
Both&   100  &   -2.4  &   0.4  &   20.5  &    5.4  &    5.3  &    5.4    \\ 
&    99  &   -2.7  &   0.4  &   18.5  &    5.4  &    5.2  &    5.1    \\ 
&    95  &   -3.3  &   0.3  &   14.9  &    5.2  &    4.7  &    4.0    \\ 
&    80  &   -3.9  &   0.3  &   10.3  &    4.6  &    4.0  &    1.3    \\\hline 
  \end{tabular}
  } 
  \hspace{2EM}
  \parbox{0.45\linewidth}{\centering {\large GOODS-N} \\
    \begin{tabular}{c|c@{ $\pm$ }l|ccc|c}
    cut  
    & $\langle\Delta\rangle$ &
    $\frac{\sigma_\Delta}{\sqrt{N}}$  
    & $\sigma_\Delta$ & $\sigma_{68}$ & $\sigma_{95}/2$ & outl. \\\hline
   100  &   11.8  &   1.1  &   48.5  &    4.2  &    9.4  &   11.9    \\ 
    99  &   11.0  &   1.1  &   46.4  &    4.1  &    8.6  &   11.3    \\ 
    95  &   10.0  &   1.1  &   44.0  &    4.0  &    7.4  &   10.7    \\ 
    80  &    9.8  &   1.2  &   43.9  &    4.1  &    7.3  &   10.5    \\\hline 
   100  &    3.2  &   0.7  &   27.8  &    3.9  &    4.9  &    7.8    \\ 
    99  &    2.1  &   0.5  &   23.1  &    3.9  &    4.7  &    7.3    \\ 
    95  &    0.8  &   0.3  &   14.1  &    3.7  &    4.3  &    6.3    \\ 
    80  &   -0.2  &   0.2  &    9.1  &    3.1  &    3.0  &    1.9    \\\hline 
   100  &    1.5  &   0.6  &   23.8  &    3.7  &    4.5  &    5.8    \\ 
    99  &    0.9  &   0.5  &   21.0  &    3.6  &    4.4  &    5.4    \\ 
    95  &   -0.1  &   0.3  &   13.1  &    3.4  &    4.0  &    4.4    \\ 
    80  &   -0.02  &   0.3  &   10.9  &    3.0  &    3.0  &    2.7    \\\hline 
   100  &    1.3  &   0.5  &   21.4  &    3.5  &    4.1  &    5.5    \\ 
    99  &    0.4  &   0.4  &   16.9  &    3.5  &    4.0  &    5.0    \\ 
    95  &    0.1  &   0.3  &   12.7  &    3.3  &    3.7  &    4.4    \\ 
    80  &   -0.3  &   0.2  &    9.3  &    2.8  &    2.7  &    2.0    \\\hline 
  \end{tabular}
  }
  \caption{
    Summary of photo-z performance results for
    unweighted, magnitude, SB, and combination priors for VVDS (left)
    and GOODS-N 
    (right).  The subsequent rows for each prior beyond the first show
    how cutting a certain fraction of the galaxies based on the
    combination goodness-of-fit of the templates and prior (i.e.\ the
    final posterior $\chi^2$ value) enables one
    to trade improved scatter for slightly 
    worse bias, eliminating a larger fraction of the outliers.
    The columns of the table are the fraction of sample remaining,
    $\langle\Delta\rangle$ and uncertainty, 
    $\sigma_\Delta$, $\sigma_\Delta$ at the 68$^{\rm th}$ percentile,
    $\sigma_\Delta$ at the 95$^{\rm th}$ percentile, and fraction of
    outliers, all expressed as percentages; we
    follow the convention and define an outlier as having $|\Delta| > 0.2$.
     }
  \label{tab:chisq_cuts}
\end{table*}

\subsection{Applications to weak lensing}

Lensing tomography refers to the use of depth information in the
source galaxies to get three-dimensional information on the lensing
mass \citep{Hu:1999ek}. By binning source galaxies in photo-z bins, the
evolution of the lensing power spectrum can be measured. 
This greatly improves the sensitivity of lensing to
dark energy in cosmological applications. The relative shift in
the amplitudes of the lensing spectra is sensitive to the 
properties of dark energy.  It depends on both distances
and the growth of structure, thus enabling tests of dark energy or
modified gravity explanations for the cosmic acceleration. 
Any errors in the bin redshifts propagate to the inferred
distances and growth factors and thus degrade the
ability to discriminate cosmological models.

The capability of lensing surveys to meet their
scientific goals will depend on our ability to characterize
the photo-z scatter, bias and fraction of outliers. 
Photometric redshifts must be calibrated with an appropriate sample of 
spectroscopic redshifts \citep{Huterer:2005ez, Ma:2005rc}.
This may be done more cheaply by using 
auto- and cross-correlations of photometric and
spectroscopic redshifts samples \citep{Schneider:2006br}, which
can also be used to estimate the redshift distribution 
for a galaxy sample where the calibration data is incomplete 
\citep[see also][]{Schneider:2006br}. 

For lensing power spectrum measurements, broad bins in photo-z are
sufficient -- bin width $\Delta z \simeq 0.2$ or larger. This makes
photo-z scatter less of an issue, since wide area surveys average over
million(s) of galaxies per bin. However residual bias in the estimated
mean redshift per bin leads to an error in cosmological parameters;
this systematic error can dominate the error budget if the photo-z's
are not well characterized and calibrated. In \citet{Huterer:2005ez,
  Ma:2005rc, 2006AAS...209.8606N} it is shown that the next generation ground
based surveys that cover $\sim 1000$ square degrees require residual
bias levels below 0.01, and our method is close to this goal in the
$0.4<z<1.3$ range; however the more ambitious surveys planned by LSST
and SNAP require levels below 0.002.  Such an exquisite control over
photo-z biases requires improvements in photo-z techniques and very
demanding spectroscopic calibration data.

Our results on the surface brightness prior show that it will be
valuable in eliminating outliers that can bias photo-z's. In
the literature outlier clipping is often performed before photo-z's 
are evaluated, but there is not a well established basis for how the
clipping can be done with real data. This is part of the reason for
why photo-z's do not perform as well on data as expected from tests on
simulated galaxy colors. With more realistic testing, we expect that
the surface brightness prior will emerge as an essential part of
photo-z measurements for lensing and other applications. Also of value
is understanding the relation to galaxy type and redshift: for lensing
it is permissible to use sub-samples that have well-behaved redshift
distributions (see \citet{Jain:2006gi} for an application of
this idea to ``color tomography''). 

Lensing conserves surface brightness, but it can introduce
magnification bias in the apparent magnitude. Hence using magnitude
priors can cause subtle biases in lensing measurements; e.g., behind
galaxy clusters magnitudes are brighter than in the field. Thus
photo-z's estimated with a magnitude prior would place galaxies that
are behind clusters at lower redshift than galaxies that are in the
field, and the effect would be more pronounced for more massive
clusters compared to smaller ones, thus biasing cosmological
inferences from lensing measurements.  While the bias is expected to
be small, at the percent level, it does argue in favor of surface
brightness priors for precision lensing measurements.

While we have discussed weak lensing induced shear correlations,
similar conclusions apply to other lensing applications such as galaxy-shear
cross-correlations (known as galaxy-galaxy lensing and discussed recently by
\citet{Mandelbaum:2007dp}). For cluster lensing, i.e. making maps of the
projected mass distribution using weak and strong lensing, the scatter in the
photo-z relation is more important since only the source galaxies
in the cluster field provide useful lensing information. 





\section{Discussion}
\label{sec:discussion}

We have shown that using the surface brightness-redshift relation, we
can improve template-based photo-z results for both ground and space
data.  Our main result, presented in Sec.~\ref{sec:results}, shows how
the SB prior is able to help eliminate much of the color-redshift
degeneracy that is present for many galaxies in the GOODS and VVDS
samples. When compared to a standard unweighted redshift estimator, SB
priors reduce the fraction of outliers in the photo-z relation from
12--16\% to $\sim$6\%. This results in a decrease in the scatter in
the photo-z relation (as defined by the 95$^{\rm th}$ percentile of
the distribution) by a factor of two or more, while the bias improves
from a value of $\Delta \sim 0.1$ to $\Delta \sim 0.05$; almost all
redshift and type bins improve in the VVDS data, whereas the GOODS
data improves likewise for all types but with an especially large
effect at redshifts $z<0.6$.

At redshifts $z>1.3$ the limitations of the passbands used in the VVDS
give rise to a bias that becomes more negative with increasing
redshift.  The utility of the SB prior is dependent on seeing and the
size measures provided; in the VVDS data, the sole public size measure
(ellipse area measured by SExtractor) is not precise enough to
optimally measure the surface brightness-redshift relation. We expect
that more sophisticated techniques, such as fitting a light profile,
would reduce the sensitivity of the SB prior to seeing; in particular,
we require a size estimator that minimizes the scatter in the SB($z$)
relation.  If a ground-based survey can precisely measure the angular
area of galaxies and achieve a seeing of 1" or less,
with a representative calibration sample that has the same
sensitivity, then such a ground-based survey should be able to do
almost as well as one from space with respect to constraining
photo-z's using the SB.

Since our current application utilizes spectroscopic data to calibrate
and test the SB prior, we are necessarily restricted to studying
bright galaxies.  Such a regime naturally favors a magnitude prior; as
Fig.~\ref{fig:magprior} shows, the bright galaxy population is almost
exclusively at low redshift.  As the galaxies become fainter, the
magnitude prior will have less and less predictive power;
unfortunately, since most galaxies are faint, this is precisely the
galaxy population we are most interested in obtaining photometric
redshifts for.  We do not expect the SB prior to suffer from this
effect as much; if the sizes of galaxies are measured carefully so as
to minimize the scatter in the SB($z$) relation, then the SB prior
should retain its statistical power for faint galaxies as well.

Based on these results we expect, therefore, that the use of SB in
constraining photometric redshifts (whether using a template based
approach, neural networks, or other empirical relations) can
substantially improve the robustness of the application of photo-z's
in cosmology. The details of the application and its impact will be
application dependent. In some applications, such as WL
tomography \citep{Huterer:2005ez, Ma:2005rc}, it is important for the
measured redshifts to be as unbiased as possible, while others, such as
Baryon Oscillations, are more sensitive to the scatter in the
measurement. Subsequent papers will address the question of
optimal measures for size from ground-based data and the resulting
improvements in the SB prior.

\section*{Acknowledgments} 

We are very grateful to Mike Jarvis for helpful discussions and coding
contributions.  HFS especially thanks Masahiro Takada for his
hospitality and support, access to Subaru data, and invaluable
discussions.  We thank Mariangela Bernardi, David Wittman, and Mitch
Struble for their comments, and Ofer Lahav, Felipe Abdalla, Huan Lin,
Gary Bernstein, and Ravi Sheth for helpful discussions.  This work is
supported in part by NSF grant AST-0607667, the Department of Energy,
and the Research Corporation.  AJC acknowledges partial support from
NSF ITR 0312498 and NSF ITR 0121671 and NASA grants NNX07-AH07G and
STSCI grant AR-1046.04

\bibliographystyle{mn2e}
\bibliography{photoz}

\end{document}